\documentclass[apj]{emulateapj}

\usepackage{textcomp}
\usepackage{bm,color}
\usepackage{verbatim}
\usepackage{amsmath}
\usepackage{epstopdf}
\usepackage{ulem}
\usepackage{graphicx}

\def\apjl{ApJ}%
\def\apjs{ApJS}%
\def\ao{Appl.~Opt.}%
\def\aap{A\&A}%
\def\simlt{\lower.5ex\hbox{$\; \buildrel < \over \sim \;$}}
\def\simgt{\lower.5ex\hbox{$\; \buildrel > \over \sim \;$}}
\def\simgtalt{\lower.5ex\hbox{$\buildrel > \over \sim \;$}}

\def\bd#1{{\bf #1}}
\def\com#1{{#1}}

\def\eref#1{(\ref{#1})}

\usepackage{amsfonts}

\begin{document}

\title{\com{Modeling} atmospheric \com{emission} for CMB ground-based observations}

\author{
J.~Errard\altaffilmark{30,3},
P.A.R.~Ade\altaffilmark{27},
Y.~Akiba\altaffilmark{16},
K.~Arnold\altaffilmark{14},
M.~Atlas\altaffilmark{14},
C.~Baccigalupi\altaffilmark{17},
D.~Barron\altaffilmark{14},
D.~Boettger\altaffilmark{5},
J.~Borrill\altaffilmark{3,30},
S.~Chapman\altaffilmark{9},
Y.~Chinone\altaffilmark{16,13},
A.~Cukierman\altaffilmark{13},
J.~Delabrouille\altaffilmark{1},
M.~Dobbs\altaffilmark{24},
A.~Ducout\altaffilmark{11},
T.~Elleflot\altaffilmark{14},
G.~Fabbian\altaffilmark{17},
C.~Feng\altaffilmark{32},
S.~Feeney\altaffilmark{11},
A.~Gilbert\altaffilmark{24},
N.~Goeckner-Wald\altaffilmark{13},
N.W.~Halverson\altaffilmark{2,6,15},
M.~Hasegawa\altaffilmark{16,31},
K.~Hattori\altaffilmark{16},
M.~Hazumi\altaffilmark{16,31,19},
C.~Hill\altaffilmark{13},
W.L.~Holzapfel\altaffilmark{13},
Y.~Hori\altaffilmark{13},
Y.~Inoue\altaffilmark{16},
G.C.~Jaehnig\altaffilmark{2,15},
A.H.~Jaffe\altaffilmark{11},
O.~Jeong\altaffilmark{13},
N.~Katayama\altaffilmark{19},
J.~Kaufman\altaffilmark{14},
B.~Keating\altaffilmark{14},
Z.~Kermish\altaffilmark{12},
R.~Keskitalo\altaffilmark{3},
T.~Kisner\altaffilmark{3,30},
M.~Le~Jeune\altaffilmark{1},
A.T.~Lee\altaffilmark{13,25},
E.M.~Leitch\altaffilmark{4,18},
D.~Leon\altaffilmark{14},
E.~Linder\altaffilmark{25},
F.~Matsuda\altaffilmark{14},
T.~Matsumura\altaffilmark{33},
N.J.~Miller\altaffilmark{21},
M.J.~Myers\altaffilmark{13},
M.~Navaroli\altaffilmark{14},
H.~Nishino\altaffilmark{19},
T.~Okamura\altaffilmark{16},
H.~Paar\altaffilmark{14},
J.~Peloton\altaffilmark{1},
D.~Poletti\altaffilmark{1},
G.~Puglisi\altaffilmark{17}, 
G.~Rebeiz\altaffilmark{7},
C.L.~Reichardt\altaffilmark{29},
P.L.~Richards\altaffilmark{13},
C.~Ross\altaffilmark{9},
K.M.~Rotermund\altaffilmark{9},
D.E.~Schenck\altaffilmark{2,6},
B.D.~Sherwin\altaffilmark{13,20},
P.~Siritanasak\altaffilmark{14},
G.~Smecher\altaffilmark{24},
N.~Stebor\altaffilmark{14},
B.~Steinbach\altaffilmark{13},
R.~Stompor\altaffilmark{1},
A.~Suzuki\altaffilmark{13},
O.~Tajima\altaffilmark{16},
S.~Takakura\altaffilmark{22,16},
A.~Tikhomirov\altaffilmark{9},
T.~Tomaru\altaffilmark{16},
N.~Whitehorn\altaffilmark{13},
B.~Wilson\altaffilmark{14},
A.~Yadav\altaffilmark{14},
O.~Zahn\altaffilmark{25}}

\begin{abstract}
Atmosphere is one of the most important noise sources for ground-based cosmic microwave background (CMB) experiments. By increasing optical loading on the detectors, it amplifies their effective noise, while its fluctuations introduce spatial and temporal correlations between detected signals.
We present a physically motivated 3d-model of the atmosphere total intensity emission in the millimeter and sub-millimeter wavelengths. We derive \com{a new} analytical estimate for the correlation between detectors time-ordered data as a function of the instrument and survey design, as well as several atmospheric parameters such as wind, relative humidity, temperature and turbulence characteristics. 
Using \com{an original} numerical computation, we examine the effect of each physical parameter on the correlations in the time series of a given experiment. We then use a parametric-likelihood approach to validate the modeling and estimate atmosphere parameters from the \textsc{polarbear-i} project first season data set. \com{We derive a new 1.0\% upper limit on the linear polarization fraction of atmospheric emission. We also compare our results to previous studies and weather station measurements}. The proposed model can be used for realistic simulations of future ground-based CMB observations.
\end{abstract}

\maketitle

\section{Introduction}

Observations of the cosmic microwave background (CMB) are a unique probe of the fundamental physics at work in the early universe. Ongoing and upcoming high sensitivity observations of CMB temperature and polarization anisotropies provide constraints on the properties of the very early Universe, for example through observations of CMB primordial polarization from large to intermediate angular scales. 
In addition, through the precise characterization of the impact of gravitational lensing on smaller scales, the analysis of CMB polarization brings information about large scale structures, and corresponding constraints on neutrinos masses and species and on the dark energy, e.g., \citet{2013arXiv1303.5062P, 2013arXiv1301.1037D}.

Ground-based experimental efforts have been playing and wil continue to play a prominent role in the exploitation of this potential
 (e.g. \cite{2014arXiv1403.2369T, 2014arXiv1403.3985B, 2014JCAP...10..007N,2014arXiv1411.1042C, 2013arXiv1309.5383A, 2014ApJ...788..138W}).
 As compared to space-borne instruments such as WMAP\footnote{http://map.gsfc.nasa.gov/} or Planck\footnote{http://www.esa.int/Our\_Activities/\\ Space\_Science/Planck}, or planned next-generation space mission such as COrE\footnote{http://www.core-mission.org/}, LiteBIRD\footnote{http://litebird.jp/eng/} and PIXIE (\cite{2011arXiv1105.2044K}), ground-based experiments can deploy larger primary reflectors and therefore reach higher angular resolution, but must deal with residual atmospheric effects which can be an important limitation to their ultimate performance.
Emission from water vapor and dioxygen molecules dominates atmospheric emission at millimeter and sub-millimeter wavelengths. This effect can be significantly minimized by observing in so-called atmospheric windows, where emission levels are greatly reduced. However, even this residual emission inevitably increases the optical power incident on the detectors and therefore their  (photon) noise level, e.g., \cite{2010PhDT.......176A}. In addition, the inhomogeneous distribution of water vapor molecules, driven by complex mechanisms, which depend on the properties of the atmosphere above a given observation site, results in both temporal and spatial variations of the received optical power. The amplitude and correlation of atmospheric fluctuations depend on both the scanning strategy and the properties of the atmosphere at the time of observation, such as  the wind direction and speed.
If treated as an additional noise-like component, this atmospheric contamination results in an additional colored and spatially correlated signal in the time stream of any specific detector. This would potentially affect not only the effective noise level of the entire instrument, obtained as a combination of the noise levels of all its detectors, but also the rate at which this noise level decreases when the number of deployed detectors grows.


The atmospheric emission is expected to be largely unpolarized, e.g., \cite{hanany_rosenkraz, 2011MNRAS.414.3272S} and therefore appropriate hardware solutions, such as the dual polarized TES (\cite{2010PhDT.......176A}), may be used to mitigate its impact on the polarization-sensitive observations of the CMB. Nevertheless, even if such measures are implemented, the atmospheric emission will in general contribute to the detected polarized signals due to 
instrumental limitations such as instrumental polarization, imperfect half-wave plates or frequency bandpass mismatch (e.g., \cite{2008PhRvD..77h3003S, thesis_jojo}). Moreover, total intensity sensitivity is crucial for calibration, which can impact, for example, \com{the performance of component separation} based on a parametric approach, e.g. \cite{2009MNRAS.392..216S}.
Therefore,  though the focus of this work is on total intensity measurements, its conclusions are relevant for polarization-sensitive observations alike.

 Modeling of the atmospheric effects is complex. Fluctuations of the atmospheric optical depth generate emission with amplitude and scale which both depend on the observation site (dryness, air density, etc.) and on the time of observation (temperature, pressure, etc.). Moreover, wind can displace atmospheric structures, introducing hard-to-model non-stationary effects. Nonetheless, it has been shown, \cite{2000ApJ...543..787L, 2005ApJ...622.1343B}, that in the case of a given experimental setup (focal plane, scanning strategy, etc.), and with help of some simplifying assumptions useful models can be developed and implemented. These previous efforts have used the 2d frozen screen approximation to characterize observations above the south pole. In contrast, we employ a 3d model, derived \com{as a generalization of the model of \cite{1995MNRAS.272..551C}} and, \com{for the first time}, apply it to the data gathered at the Atacama Desert. The modeling proposed here requires significant computational resources,  but is more general and permits, for instance, accounting for near- and far-field beam regimes. \\ 

The paper is organized as follows. In section~\ref{sec:photon-noise}, we describe the total emission of the atmosphere, its impact on the photon noise of a ground-based detector as a function of frequency, and the typical emission law of atmospheric emission fluctuations due to water vapor inhomogeneity. In section~\ref{sec:atmosphere_physical_model}, we derive \com{a new} general expression for the auto- and cross-correlation induced by atmosphere between two detectors of a given focal plane geometry. We present in section~\ref{sec:numerical_computation} several results obtained with the model, computed using Quasi Monte Carlo algorithms, to illustrate the impact of the various parameters on the properties of the atmospheric signal.  In section~\ref{sec:atmosphere_estimation}, we compare the modeling predictions with real CMB data sets from the first season of  \textsc{polarbear-i} observations. In addition, we derive estimates of the atmosphere model parameters, and compare them with measurements taken \com{by a nearby} weather station. We also compare our results with measurements made in \cite{2000ApJ...543..787L} above the Atacama desert, and estimate an upper limit for the polarization fraction of the atmosphere.
Finally, we conclude and discuss our results in section~\ref{sec:conclusion}.
As a complement, we introduce in appendix~\ref{sec:approx_cov} a faster method to estimate an approximation of the correlated noise between detectors, encapsulated in a full binned noise covariance matrix, and to efficiently simulate realistic noise time streams.

\section{Atmospheric transmission and emission}
\label{sec:photon-noise}

The atmosphere is only partly transparent at millimeter wavelengths. Water vapor is responsible for most of the continuum absorption in the $100\,$GHz--$1\,$THz frequency range. In addition, very strong absorption occurs at frequencies within the broad oxygen absorption band around $60\pm10\,$GHz, as well as around the oxygen line at 119 GHz and the water vapor lines at 22, 183, 325, and $380\,$GHz. This leaves only four main atmospheric windows available for CMB observations from the ground: below $50\,$GHz, and around 95, 150 and $250\,$GHz. Observations are also possible at about $340\,$GHz, but only when the atmosphere is exceptionally dry, i.e., for a low Precipitable Water Vapor (PWV) value.  At higher frequencies, atmospheric opacity is incompatible with sensitive ground-based CMB observations, even from the best observing sites like the Atacama desert. 
\begin{table}[th]
\begin{center}
{\footnotesize
\begin{tabular}{|l|cccc|}
\hline 
{\bf PWV [mm]}  {\hspace {10mm}}& 0.25 & 0.5 & 1.0 & 2.0 \\
\hline 
{\bf time below} {\hspace {10mm}} &  {\hspace {1mm}} 5\% {\hspace {1mm}} &  {\hspace {1mm}} 25\% {\hspace {1mm}} &  {\hspace {1mm}} 50\% {\hspace {1mm}} &  {\hspace {1mm}} 65\% {\hspace {1mm}}  \\
\hline 
\end{tabular}
}
\end{center}
\caption{Percentage of time below the listed amount of total precipitable water vapor, for the Llano de Chajnantor in Chile.
}
\label{tab:atacama-pwv}
\end{table}

\begin{figure}[th!]
	\centering
		\includegraphics[width=\columnwidth]{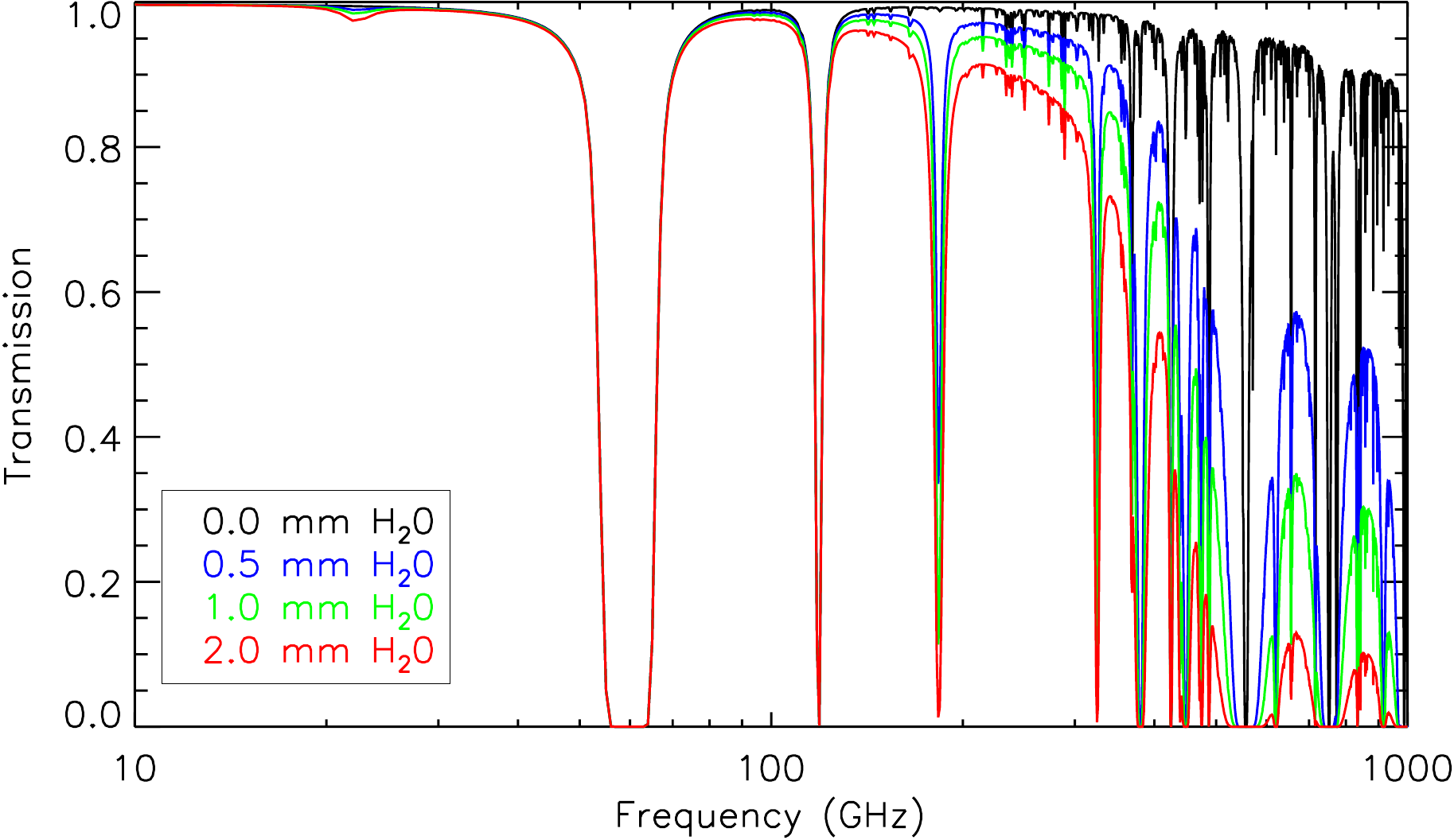}
	\caption{Atmospheric transmission from the Atacama plateau at the zenith for different amounts of precipitable water vapor. This is obtained using the ATM code, \cite{2001ITAP...49.1683P}.}
	\label{fig:transmission}
\end{figure}

Table \ref{tab:atacama-pwv}, which summarizes the percentage of time at Atacama when the PWV is below a certain amount, shows that the typical amount of water vapor is about $1$ mm, the atmospheric conditions being better about $50\%$ of the time\footnote{https://almascience.nrao.edu/\,documents-\,and-\,tools/\\overview/\,about-\,alma/atmosphere-model}. We use the Atmospheric Transmissions at Microwaves (ATM) code (\cite{2001ITAP...49.1683P}) to compute the atmospheric transmission $\mathcal T$, defined as the frequency-dependent ratio $I/I_0$ between the radiation $I$ received by a detector and the radiation $I_0$ above the atmosphere. At the zenith, from the ALMA site located at $5040$ m elevation in the Atacama desert in Chile (Llano de Chajnantor), we get the transmission curves shown in Fig.~\ref{fig:transmission}.

For different zenith angles, the optical depth $\tau$ has to be modified for the airmass $m\left( 90^\circ - {\rm el} \right)$, so that the atmospheric transmission is 
\begin{equation}
	\centering
		\mathcal T = \exp(-\tau) = \exp(-m\left( 90^\circ - {\rm el} \right) \, \tau_0).
	\label{eq:transmission_def}
\end{equation}
The airmass $m\left(90^\circ - {\rm el}\right)$ ranges from 1 at zenith (${\rm el}=90^\circ$) to about 40 at the horizon (${\rm el}=0^\circ$), and $\tau_0$ is the optical depth at the zenith. The airmass can be computed for any zenith angle using the following fitting formula by \cite{1989ApOpt..28.4735K}:
\begin{equation}
m(\theta) \propto \frac{1}{ \cos\theta + 0.50572 (96.07995-\theta\, [\deg])^{-1.6364} }.
\end{equation}
We note that for high elevations we have $m(90^\circ - {\rm el}) \simeq 1/\sin\left({\rm el}\right)$, as expected for a plane-parallel approximation.

\com{The instantaneous noise of current CMB detectors is dominated by} the statistical fluctuations in photons arriving at the detectors. Consequently instrument environment, specifically ground or space, plays a critical role.
On the one hand, for space-borne detectors, the total background in the frequency range of interest for CMB observations is the sum of the CMB blackbody emission and the emission from the optics, dominated by the emission from telescope mirrors,
\begin{equation}
I(\nu) = B_\nu(T_{\rm CMB}) + \varepsilon_{\rm tel}(\nu)B_\nu(T_{\rm tel}),
\end{equation}
where $B_\nu(T)$ is the emission of a thermal blackbody at temperature $T$. 
For a space mission such as Planck, $T_{\rm tel} \simeq 40$ K, and on the basis of the pre-launch measurements performed on the Planck reflectors (\cite{2010A&A...520A...2T}), we can model the emissivity of a two-reflector space-borne telescope as 
\begin{equation}
\varepsilon_{\rm tel}(\nu) = 0.002 \times \left( \frac{\nu}{140\,{\rm GHz}} \right)^{0.5}.
\end{equation}
On the other hand, for ground-based experiments, the temperature of the telescope and cryostat window is of the order of $280$ K and the total emissivity is typically of the order of a few percent. In addition, the emission of the atmosphere contributes to the total background with an additional term $\mathcal E(\nu)$ of
\begin{equation}
\mathcal E(\nu) = \left[1 \!-\! \mathcal T(\nu)\right] \, B_\nu(T_{\rm atm}),
\end{equation}
where $T_{\rm atm} \simeq 280$ K is the atmosphere temperature. Thus, at $150$ GHz, the total background emission is $\sim\,20$ K on the ground while it is $\sim\,1$ K in space.

In addition to this background emission, the emission of the atmosphere fluctuates as a function of time and pointing direction. A model of these fluctuations, due to inhomogeneities of its physical properties, is discussed in the next sections. 

\section{Atmosphere physical model}
\label{sec:atmosphere_physical_model}

In this section, we present the modeling of the atmospheric contamination occurring for ground-based observations, through the derivation of an analytical expression based on the auto- and cross-correlation between detectors time streams, initially proposed by \cite{1995MNRAS.272..551C}. Since the atmosphere mainly radiates non-polarized light, the detectors considered throughout the following sections are assumed to give measurements of the sky total intensity, i.e., the Stokes parameter $I$, expressed in $K$. For example, in the case of antenna-coupled TES (\cite{2010PhDT.......176A}), the considered time stream would effectively be the sum of the time-ordered data (TOD) coming from two orthogonal antennas confined within a focal plane pixel.

\subsection{Atmosphere contribution to antenna temperature}

\com{Spatial inhomogeneities} in the atmosphere emission can get imprinted in the TOD when the instrument line of sight scans across them. In addition, the wind blows structures through the beam. These inhomogeneities are due to turbulences along the line-of-sight, and are sometimes assumed to be concentrated in a two-dimensional layer: atmospheric contamination is then approximated as a screen of brightness temperature, moving with the wind (\cite{2000ApJ...543..787L, 2005ApJ...622.1343B,2010ApJ...708.1674S}). Furthermore, these fluctuations are usually assumed to be frozen as the involved turbulent processes are much smaller than the displacements imposed by the wind (\cite{1938RSPSA.164..476T}).

The model proposed by \cite{1995MNRAS.272..551C} is not restricted to a single turbulent layer, but instead sees the atmosphere as a continuum three-dimensional medium that depends on water vapor distribution, wind speed, temperature, etc. as one moves away from the telescope, along the line of sight. Our work consists in the implementation and exploitation of this $3$d modeling, hence significantly differing from previous studies.
Based on the Kolmogorov model of turbulence (\cite{1961wptm.book.....T}), the power spectrum of the fluctuations in a large three-dimensional volume is 
\begin{eqnarray}
	\centering
		P(\mathbf{k}) \propto k^{b}
	\label{eq:Pk_kolm}
\end{eqnarray}
where $k$ is the wavenumber \com{with} units $m^{-1}$, and $b =-11/3$. 
As a consequence, one can show that the contaminated time streams have power law spectra, with an index related to the Kolmogorov spectrum, typically $P(f) \propto f^{b}$, e.g., \cite{ thesis_marriage, thesis_jojo}.
Following~\cite{1995MNRAS.272..551C}, a coherence distance scale is defined over which the Kolmogorov spectrum holds --- the outer scale $L_o$ corresponds to the typical distance that satisfies that condition.

\begin{figure}
	\centering
		\includegraphics[width=\columnwidth]{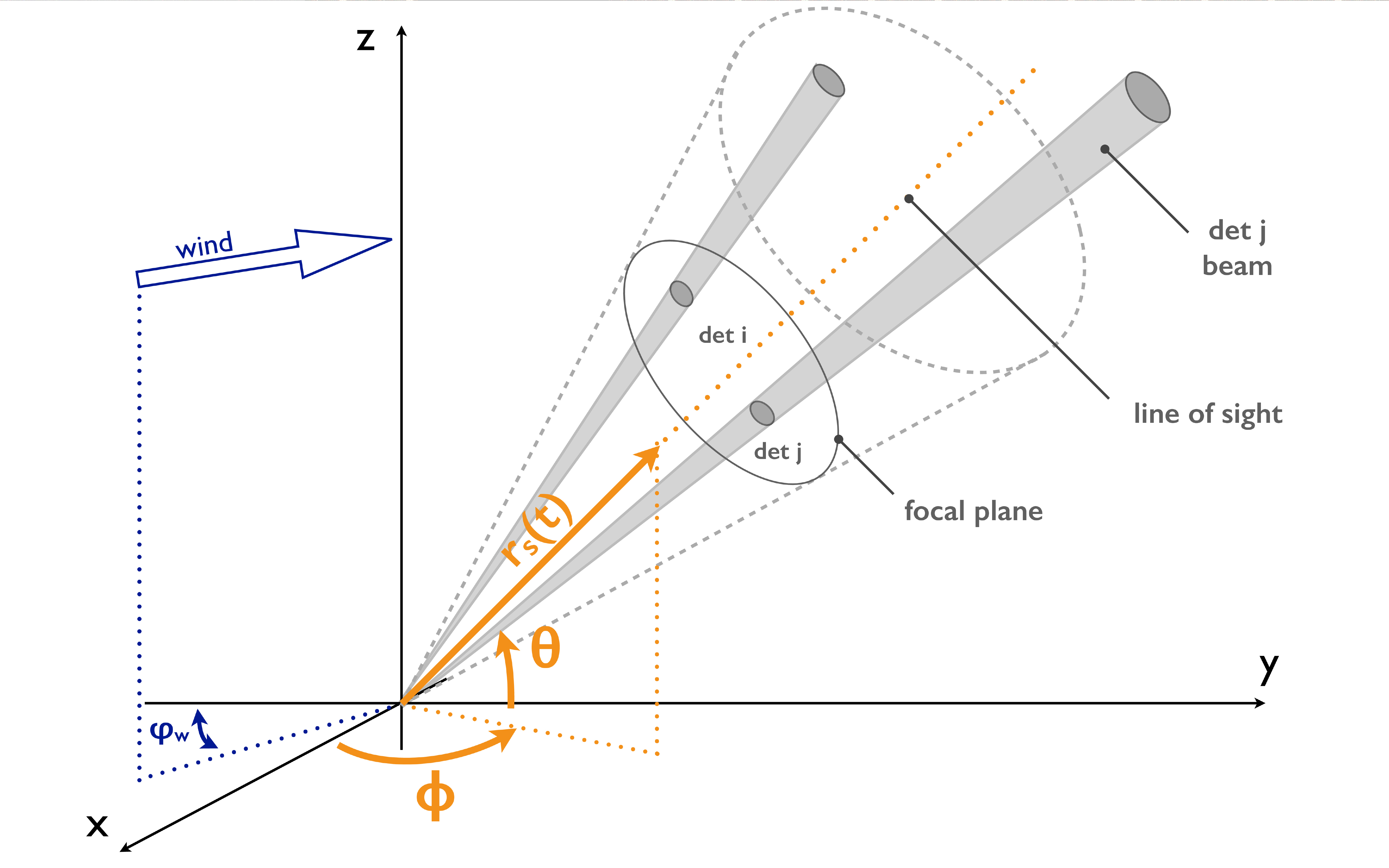}
	\caption{A detector of the focal plane observes along a vector $\bd{\hat{r}_{s}^{\ i}}$ (line of sight), defined in spherical coordinates $\phi^ i$ and $\theta^i$ (equivalently azimuth and elevation: $\theta^i = 0$ corresponds to the horizon). $\bd{\hat{r}_{s}}$ is the pointing direction of an imaginary central detector onto the focal plane. The scan strategy, $\mathbf{ss}$, can be expressed as the evolution of $\bd{\hat{r}_{s}}$ with respect to time, $\mathbf{ss}=d\bd{\hat{r}_{s}}/dt$. Wind is a vector orthogonal to the $z$-axis and is characterized by its direction $\phi_W$ and its norm $W$.}
	\label{fig:geometry}
\end{figure}

In the following reasoning, we use the geometry depicted in Fig.~\ref{fig:geometry}. An imaginary central detector of the focal plane observes along a vector $\bd{\hat{r}_{s}} = \bd{\hat{r}_{s}}(t)$, defined in spherical coordinates by $\phi$ and $\theta$ (equivalently azimuth and elevation). All other detectors are assumed to observe in directions slightly off from the central line of sight, $\bd{\hat{r}_{s}^{\ i}}(t)  = \bd{\hat{r}_{s}}(t)  + \bd{\delta r^i}$, with $ \bd{\delta r^i}$ defined by the design of the focal plane.

\begin{figure}
	\centering
		\includegraphics[width=\columnwidth]{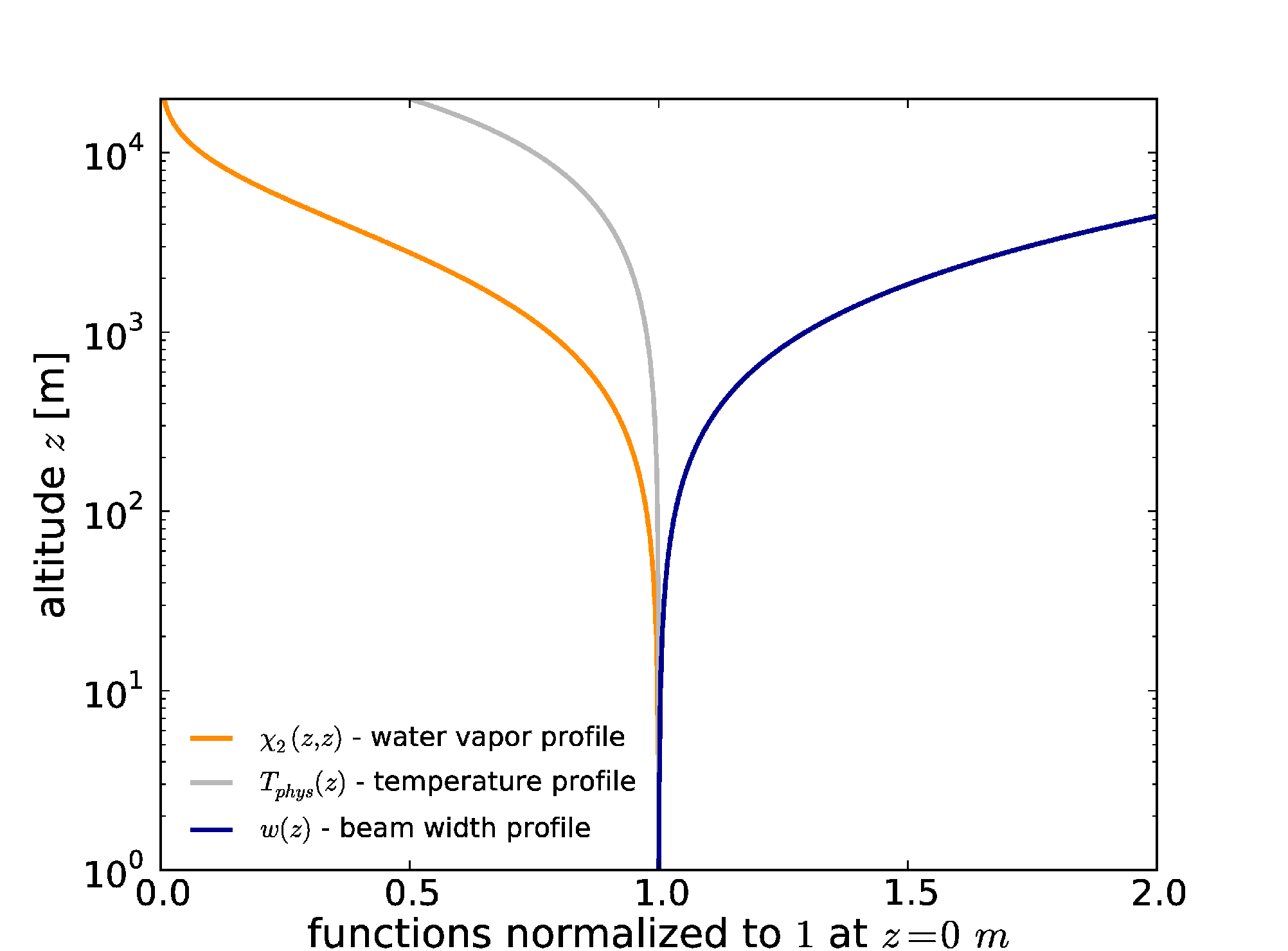}
	\caption{Illustration of three functions used in the modeling of the correlation $\mathbf{C}_{ij}^{\;tt'}$: the water vapor column, $\chi_2(z,z)$, Eq.~\eref{eq:chis}; the temperature profile, $T_{\rm phys}$, Eq.~\eref{eq:temp_adiab}; and the Gaussian beam width $w(z)$, Eq.~\eref{eq:wz}.}
	\label{fig:beam_chi2_Tphys}
\end{figure}

Each detector has an associated beam pointing in a given direction $\mathbf{\hat{r}_{s}^{\ i}}$, with an effective area $B(\mathbf{\hat{r}_{s}^{\ i}}, \mathbf{r})$ such that, for a monochromatic detector observing at wavelength $\lambda$: 
\begin{equation}
	\centering
	B(\mathbf{\hat{r}_s^i},\mathbf{r}) = \frac{2 \lambda^{2} |\mathbf{\hat{r}_s^i} \cdot \mathbf{r}|^{2}}{\pi w^{2}(\mathbf{\hat{r}_s^i} \cdot \mathbf{r})} \times \exp \left[ - \frac{2 (\mathbf{r}^{2} - (\mathbf{\hat{r}_s^i} \cdot \mathbf{r})^{2})}{w^{2}(\mathbf{\hat{r}_s^i} \cdot \mathbf{r})} \right]
	\label{eq:beff}
\end{equation}
 where
\begin{equation}
	\centering
	w(\mathbf{\hat{r}_s^i} \cdot \mathbf{r}) = w_{0} \sqrt{1 + \left( \frac{\lambda\, \mathbf{\hat{r}_s^i} \cdot \mathbf{r}}{\pi \, w_{0}^{2}} \right)^{2}}
	\label{eq:wz}
\end{equation}
with $w_{0}$ is the beam waist given by 
\begin{eqnarray}
	\centering
		w_0 \equiv \frac{ \lambda }{  \pi\, \theta_{b} },
	\label{eq:w0_def}
\end{eqnarray}
$\theta_{b}$ being the beam opening angle in radians. As expected, for large distances from the telescope, i.e., $|\mathbf{\hat{r}_s^i} \cdot \mathbf{r}| \gg w_0$, $w(\mathbf{\hat{r}_s^i} \cdot \mathbf{r})$ has a slant asymptotic with slope $w_{0}$, leading to $B(\mathbf{\hat{r}_s^i},\mathbf{r}) \propto \theta_{b}^{\; 2}$.
It is  computationally expensive to consider a realistic behavior for the beam effective area, and this was usually neglected in previous works (\cite{2000ApJ...543..787L, 2005ApJ...622.1343B,2010ApJ...708.1674S}). Fig.~\ref{fig:beam_chi2_Tphys} shows the behavior of the Gaussian beam width $w(z)$ given in Eq.~\eref{eq:wz}, compared to two other important functions used in this modeling and introduced below, the water vapor density (Eq.~\eref{eq:chis}) and the temperature profile (Eq.~\eref{eq:temp_adiab}).  

Given the expression for $B(\mathbf{\hat{r}_s^i},\mathbf{r})$, the contribution $dT_{\rm ant}$ to the antenna temperature of a small element $dV$ of atmosphere located at a distance $\mathbf{r}$ is proportional to the effective area of the telescope beam as seen from that point, i.e.,
\begin{equation}
	\centering
	dT_{\rm ant}(\mathbf{r}) = \frac{1}{\lambda^2} B(\mathbf{\hat{r}_s^i},\mathbf{r}) \, \alpha(\mathbf{r}, \lambda) \, T_{\rm phys}(\mathbf{{r}}) \, \frac{dV}{r^{2}}
	\label{eq:ant_temp}
\end{equation}
where the transmission $\mathcal{T}(\lambda)$ mentioned in Eq.~\eref{eq:transmission_def}  is related to the absorption $\alpha(\mathbf{r}, \lambda)$, expressed in $m^{-1}$, by
\begin{eqnarray}
	\centering
		 \mathcal{T}(\lambda) = e^{-\tau(\lambda)} = e^{-\int_0^r \alpha(r',\lambda) dr' }.
\end{eqnarray}
$T_{\rm phys}(\mathbf{{r}})$ in Eq.~\eref{eq:ant_temp} is the physical temperature of the given volume of atmosphere. It is convenient to assume \com{an adiabatic atmosphere}, so that \com{$T_{\rm phys}$} depends linearly on altitude $z\equiv\mathbf{r}\cdot\mathbf{\hat{z}}$,
\begin{eqnarray}
	\centering
	T_{\rm phys}(\mathbf{r}) &=& T_{\rm phys}(z) \nonumber\\ &=& T_{\rm ground} \left( 1 - \frac{z}{z_{\rm atm}} \right)
	\label{eq:temp_adiab}
\end{eqnarray}
with $T_{\rm ground} \sim 280$ K the temperature at the ground level and $z_{\rm atm}Ê\sim 10^4$ m a typical height which depends on the observation site. Note that Eq.~\eref{eq:temp_adiab} is not expected to be correct \com{above the tropopause, generally located between $10$ and $20$ km, e.g.~\cite{1993IJCli..13..461H}.} \com{We keep this assumption through the paper as our results do not depend significantly on $z_{\rm atm}$, the atmospheric emission being also weighted by the water vapor density profile. This latter decreases much more rapidly than $T_{\rm phys}$ with altitude as explained in the following paragraph.}

\subsection{Analytical expression for the auto- and cross-correlation between detectors}

By integrating Eq.~\eref{eq:ant_temp} over $\mathbf{r}$, the total antenna temperature $T_{\rm ant}$ as measured by a detector $i$ can be written as
\begin{eqnarray}
	\centering
	T^{\ i}_{ant} &\equiv&  T_{\rm ant}(\mathbf{\hat{r}^{\ i}_s}(t)) \nonumber \\
	&=& \frac{1}{\lambda^2} \int{\frac{d\mathbf{r}}{r^2} \;B(\mathbf{\hat{r}^{\ i}_s}(t),\mathbf{r}) \,\alpha(\mathbf{{r}}) \, T_{\rm phys}(\mathbf{r})}.
	\label{eq:Tant_i_def}
\end{eqnarray}

More generally, using Eq.~\eref{eq:Tant_i_def}, the correlation between two samples $t$ and $t'$, measured by two given detectors $i$ and $j$, is defined as $\langle T^{\ i}_{\rm ant}(t)T^{\ j}_{\rm ant}(t')\rangle$, expressed as:
\begin{eqnarray}
	\centering
		&& \mathbf{C}_{ij}^{\; t t'} \equiv \langle T_{\rm ant}^{\ i}(t)T_{\rm ant}^{\ j}(t') \rangle \equiv  \langle T_{\rm ant}(\mathbf{\hat{r}_{s}^{\ i}}(t))T_{\rm ant}(\mathbf{\hat{r}_s^{\ j}}(t'))\rangle = \nonumber\\ && \hspace{1.5cm} \frac{1}{\lambda^4}\int \frac{d\mathbf{r}}{r^2}\int \frac{d\mathbf{r'}}{r^2} \;B(\mathbf{\hat{r}_s^{\ i}}(t),\mathbf{r})B(\mathbf{\hat{r}_s^{\ j}}(t'),\mathbf{r'}) \nonumber\\ &&  \hspace{3cm}\times\langle\alpha(\mathbf{r})\alpha(\mathbf{r'})\rangle \, T_{\rm phys}(\mathbf{r})T_{\rm phys}(\mathbf{r'})
	\label{eq:auto2}
\end{eqnarray}
where the average  $\langle\,\cdot\,\rangle$ is taken over realizations of the sky. As in~\cite{1995MNRAS.272..551C}, we reduce the correlation $\langle\,\cdot\,\rangle$ between two given points  $\mathbf{r}$ and $\mathbf{r'}$ in the atmosphere to the $ \langle \alpha(\mathbf{r})\alpha(\mathbf{r'})\rangle$ term. Note also that we neglected the wavelengths in $\langle\alpha(\mathbf{r}, \lambda)\alpha(\mathbf{r'}, \lambda')\rangle$. We will limit ourselves to the monochromatic case in this work but will mention the  $\lambda \neq \lambda'$ possibility in section~\ref{sec:conclusion}.

Quantities in Eq.~\eref{eq:auto2} depend on the atmosphere properties (hidden in  the $\langle\alpha(\mathbf{r})\alpha(\mathbf{r'})\rangle \times T_{\rm phys}(\mathbf{r})T_{\rm phys}(\mathbf{r'})$ term) and on the experimental design and the operation of the telescope (included in the $B(\mathbf{\hat{r}_s^{\ i}}(t),\mathbf{r})B(\mathbf{\hat{r}_s^{\ j}}(t'),\mathbf{r'})$ term). At this point, if no wind is assumed, the only time dependence in Eq.~\eref{eq:auto2} is encoded in the scanning strategy i.e., $\mathbf{\hat{r}_s}(t)$. 

\begin{figure}
	\centering
		\includegraphics[width=\columnwidth]{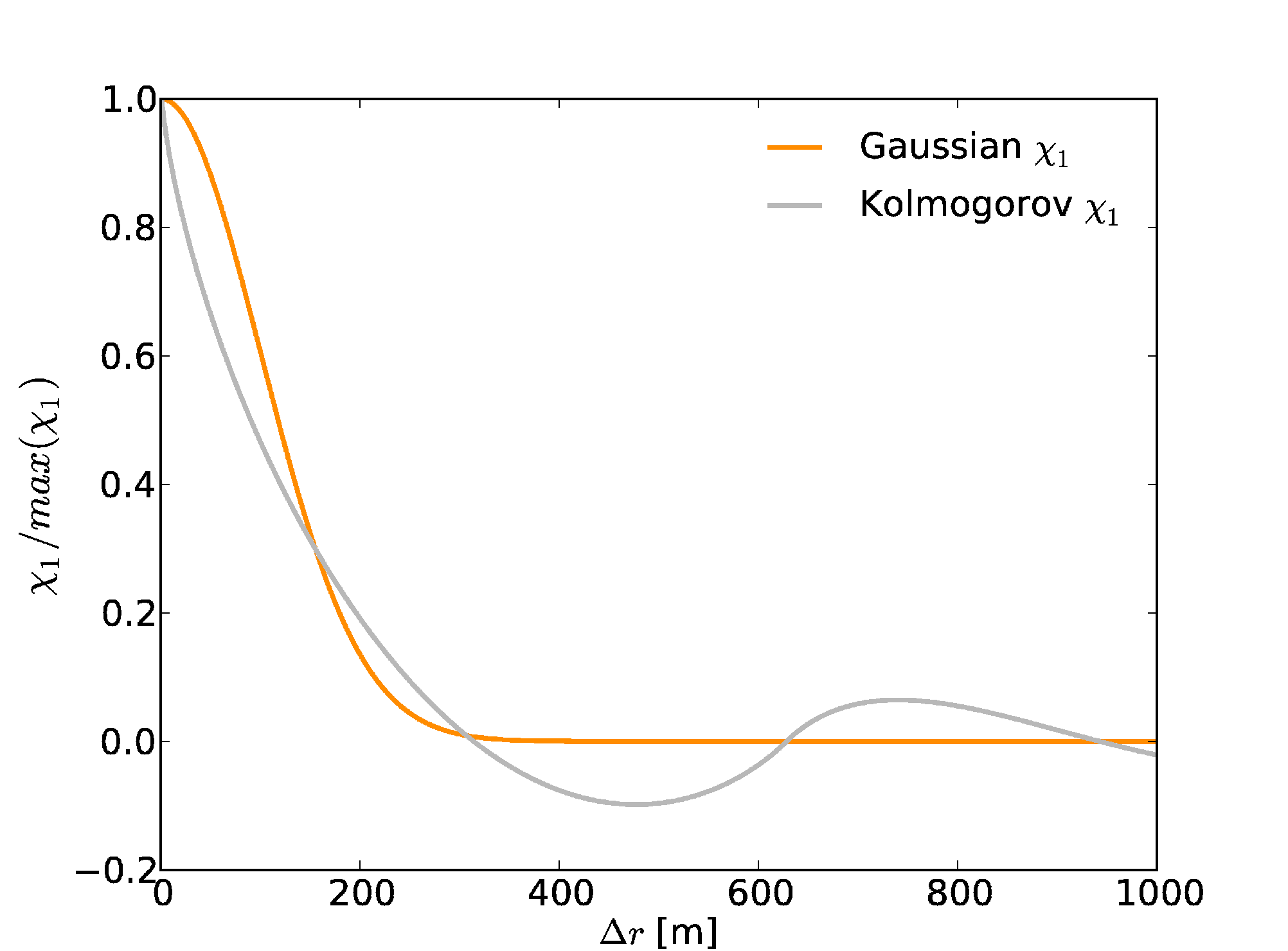}
	\caption{Comparison of $\chi^{\rm Kolm.}_{1}(\Delta r)$ (gray) given by Eq.~\eref{eq:kolmogorov_expression} and $\chi_{1}(\Delta r)$ (orange), given in Eq.~\eref{eq:chinos}, normalized to $1$ for $\Delta r = 0$ m. Here we set $L_o = 100$ m, $\kappa_{\rm min} = (L_o)^{-1} = 0.01$ m$^{-1}$ and $\kappa_{\rm max} = 100$ m$^{-1}$.}
	\label{fig:kolm_vs_gauss}
\end{figure}

To derive an analytic expression for the correlation term, $ \langle \alpha(\mathbf{r})\alpha(\mathbf{r'})\rangle$, we follow \cite{1995MNRAS.272..551C} and assume that
\begin{equation}
	\centering
		\langle\alpha(\mathbf{r})\alpha(\mathbf{r'})\rangle = \chi_{1}(| \mathbf{r} - \mathbf{r'} |) \, \chi_{2}(z, z')
	\label{eq:correlation_decomposition}
\end{equation}
where $\chi_{1}$ is the turbulent part of the correlation, effective for lengths satisfying $|\mathbf{r}-\mathbf{r'}| \leq  L_o$. In this work, as in \cite{1995MNRAS.272..551C}, $\chi_1$ is approximated by a Gaussian, i.e.,
\begin{eqnarray}
	\centering
		\chi_{1}(\Delta r) &=& \chi_1^0 \exp\left( -\; \frac{ \Delta r ^{2}}{2\,L_{o}^{2}} \right),
	\label{eq:chinos}
\end{eqnarray}
where $\Delta r \equiv |\mathbf{r}-\mathbf{r'}|$. As depicted in Fig.~\ref{fig:kolm_vs_gauss}, this is nearly equivalent to Kolmogorov correlation, given by
\begin{equation}
	\centering
		\chi^{\rm Kolm.}_{1}(\Delta r) \propto \frac{1}{\Delta r}\int_{\kappa_{\rm min}}^{\kappa_{\rm max}}{\kappa^{-11/3}{\rm sin}\left(\kappa\;\Delta r\right)\kappa d\kappa}
	\label{eq:kolmogorov_expression}
\end{equation}
where $\kappa_{\rm min}$ and $\kappa_{\rm max}$ are proportional to the inverse of typical sizes of the turbulences, respectively their outer (energy provision) and inner (energy dissipation) scales. In the case depicted in Fig.~\ref{fig:kolm_vs_gauss}, \com{the relative difference between the integration of Eqs.~\eref{eq:chinos} and \eref{eq:kolmogorov_expression} over $\Delta r$ is about 10\%.} In the following, we will use the Gaussian correlation given in Eq.~\eref{eq:chinos} for computational efficiency. \com{While the quantitative change due to the Gaussian approximation is less than $10\%$ for $\mathbf{C}_{ij}^{\; t t'}$, the Gaussian shape for $\chi_1$ drives the physical interpretation of the typical scale $L_o$}.

The second term in Eq.~\eref{eq:correlation_decomposition}, $\chi_{2}$, only depends on the altitude and can be interpreted as the water vapor distribution. This latter is assumed to be an exponential decreasing with altitude $z$,
\begin{eqnarray}
	\centering
		\chi_{2}(z, z') &=& \chi_2^0 \exp\left(-\frac{z + z'}{2\,z_0} \right),
	\label{eq:chis}
\end{eqnarray}
which seems to be a reasonable approximation as indicated by some observations made above the Chilean desert of Atacama (\cite{2001PASP..113..803G}). 
For illustration, Fig.~\ref{fig:beam_chi2_Tphys} shows $\chi_2(z,z)$ compared to the beam width $w(z)$ and the temperature profile $T_{\rm phys}(z)$. The model used for $\chi_2$ naturally depends on the site of observation. For example, water vapor density column at South Pole falls off quite significantly at altitudes outside a $300$-$2,000$ m layer (\cite{2005ApJ...622.1343B}). We come back to this difference in appendix~\ref{sec:3d_and_2d_discussion}.

We assume in this work that wind behaves such that it does not mix atmosphere between different altitudes. Wind is thought to displace atmospheric structures, in particular the fluctuations encoded in the $ \langle \alpha(\mathbf{r})\alpha(\mathbf{r'})\rangle$ correlation term. We suppose that wind affects the $\chi_{1}(\Delta r)$ term of the correlation, Eq.~\eref{eq:chinos}, and leads to a new time dependence in this latter $\chi_{1}(\mathbf{\Delta r}) \rightarrow \chi_{1}(\mathbf{\Delta r'})$ such as
\begin{eqnarray}
	\centering
		\mathbf{\Delta r'} \equiv \mathbf{\Delta r'}(\Delta t) = \mathbf{\Delta r - \mathbf{W}}\Delta t
	\label{eq:chinos_Wind}
\end{eqnarray}
with $\Delta t \equiv \left| t'-t \right|$ and where $\mathbf{W}$ is a vector describing wind direction and amplitude.  $\mathbf{W}$ is supposed to not depend on the altitude $z$ as in \cite{1995MNRAS.272..551C} or \cite{2000ApJ...543..787L}, i.e.,
\begin{eqnarray}
	\centering
		\mathbf{W} = W_x\,\mathbf{\hat{x}} + W_y\,\mathbf{\hat{y}} \equiv W\mathbf{\hat{w}},
	\label{eq:W_simple}
\end{eqnarray}
where $\mathbf{\hat{w}}$ corresponds to the unitary vector along the direction of the wind.
It is possible to refine Eqs.~(\ref{eq:chinos_Wind},\ref{eq:W_simple}) and include a dependence of $\mathbf{W}$ on the altitude. 
We do not explore this possibility in this work and rather focus on a simple effective wind, described by a constant wind norm $W = | \mathbf{W} |\ [$m s$^{-1}]$ and a single direction $\phi_W\ [\deg]$. This approximation turns out to reasonably describe real observations, as detailed in section~\ref{sec:atmosphere_estimation}.

Amplitude terms in Eqs.~\eref{eq:chinos} and~\eref{eq:chis} \com{ensure that the dimension} of $\mathbf{C}_{ij}^{\; tt'}$ is $K^2$. In this work, we set $\chi_1^0=1.0$ and $\chi_2^0 = 1.0$ m$^{-2}$. The global amplitude of the correlation is parametrized by an effective ground temperature, $T_0$, expressed in $K_{\rm eff}$. 
Note that measurements of the true ground temperature $T_{\rm ground}$, as measured by a weather station, and the estimation of $T_0$ from $\mathbf{C}_{ij}^{\; tt'}$, cf. section~\ref{sec:atmosphere_estimation}, can give an estimate for the conversion factor between $K_{\rm eff}$ and $K$.

\subsection{Comparing the 3d and 2d approaches}
\label{ssec:3d_2d_comparison_introduction}

Previous studies such as \cite{2000ApJ...543..787L, 2005ApJ...622.1343B,2010ApJ...708.1674S} used the approximation of a 2d frozen screen. Appendix~\ref{sec:3d_and_2d_discussion} shows that the 3d modeling introduced above is general enough to reproduce the 2d screen predictions.  In fact, parameters can be tuned so that the angular correlations from both approaches are nearly equivalent on small angular scales. However, this would assume a stepped function for the water vapor column, which is not supported by observations above the Atacama desert (\cite{2001PASP..113..803G}). Unlike measurements from south pole, estimated parameters from a 2d screen approach might be difficult to interpret: in particular, the typical altitude $h$ and thickness $\Delta h$ of the turbulent layer would not have clear physical meanings in the case of distributed atmospheric turbulences.

More generally, note that there are two significant reasons for using the 3d formalism. First, the beams are overlapping where much of the emission is occurring, and this can be the case for large aperture instruments. Second, the scale length can be smaller than the thickness of the layer where the atmosphere is emitting.

\com{Despite the caveats} mentioned above, we will compare the effective brightness of the atmospheric fluctuations, as estimated from both approaches. This comparison is reasonable since the observed brightness of atmospheric fluctuations, $ B_\nu^2$, should not depend on the assumed modeling. At small angular scales, identifying Eqs.~\eref{eq:C_theta_bussmann} and \eref{eq:comparison_2d_3d} leads in particular to $T_0^2 \propto B_\nu^2$, with a coefficient of proportionality depending on $L_o$, $h$, $\Delta h$, the angular scale $\gamma$ and $z_{\rm atm}$. In section \ref{sec:atmosphere_estimation}, we will have estimates for $T_0$ and $L_o$ derived from real observational data.  Although significant uncertainties on $h$ and $\Delta h$ above the Atacama desert make any quantitative comparison between the 2d and 3d approaches quite risky, estimates of atmosphere brightness derived from both modelings are presented in section~\ref{ssec:comp_2d_screen}.

\section{Implementation and an Illustrative Example}
\label{sec:numerical_computation}

\begin{figure*}[htb!]
	\centering
		\includegraphics[width=17cm]{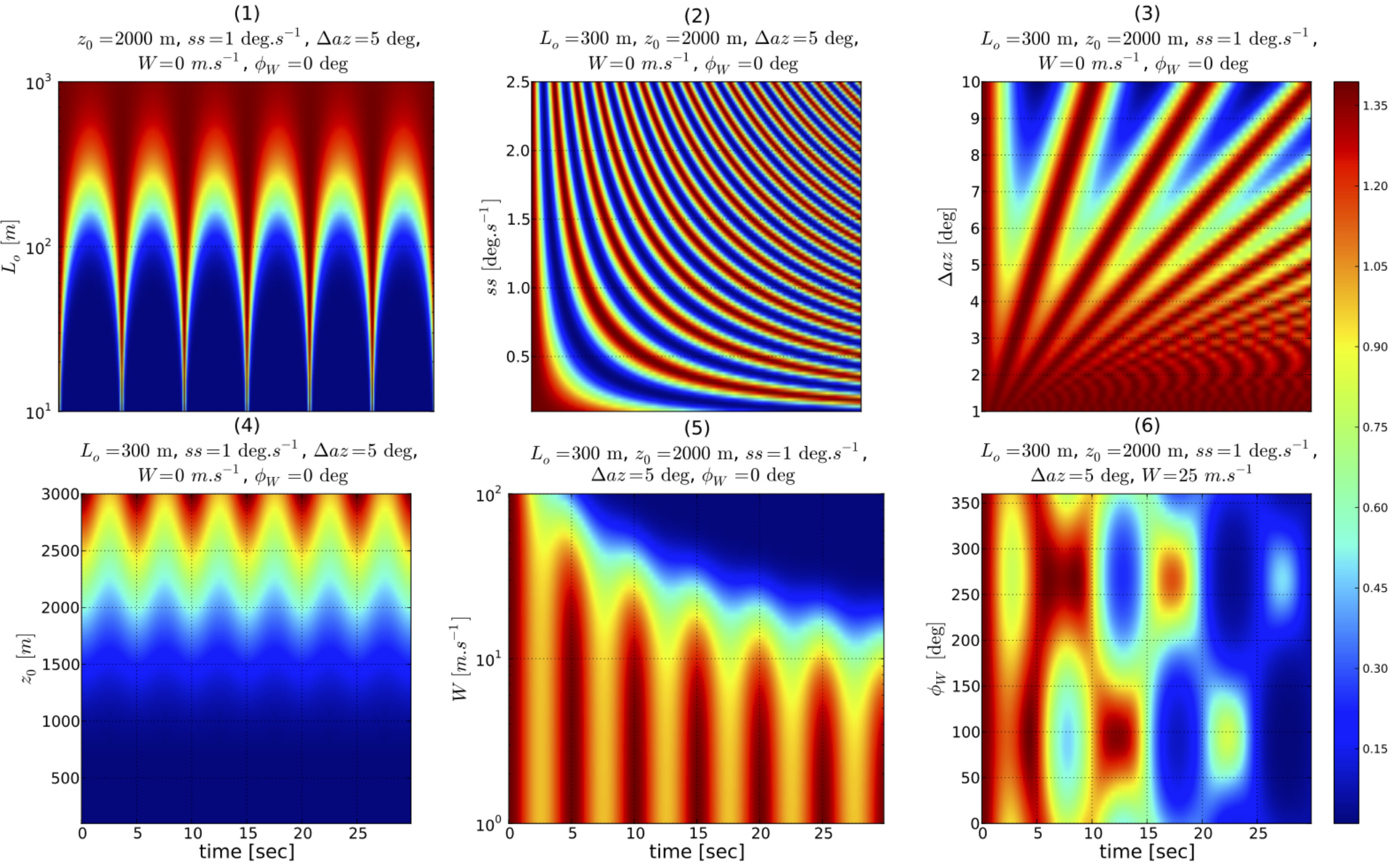}
	\caption{ Normalized auto-correlation of a detector as a function of time, i.e., $C_{00}^{\ 0t'}/  C_{00}^{\ 00,\, fid} $ as defined in Eq.~\eref{eq:auto2}, for various atmosphere physical parameters. The fiducial parameters are chosen as $\{$ $L_o=300$ m, $W=0.00$ m s$^{-1}$, $\phi_W = 0.00 \,\deg$, $\Delta az = 5.00\,\deg$, $ss=1.00\,\deg\,s^{-1}$ $\}$. The observing experiment is assumed to scan the sky in azimuth (here 6 back-and-forth movements in 30 s), at a given elevation. A detailed description of each panel can be found in the main text.}
\label{fig:Loxssxt}
\end{figure*}

In this section, we illustrate the 3d model introduced in section~\ref{sec:atmosphere_physical_model} by computing elements of $\mathbf{C}_{ij}^{\; tt'}$, Eq.~\eref{eq:auto2}, for various atmosphere conditions, as parametrized by the atmospheric fluctuations typical scale $L_o$, the wind $\mathbf{W}$, the typical altitudes $z_0$ and $z_{\rm atm}$, as well as the effective ground level temperature $T_0$. 
Contrary to \cite{1995MNRAS.272..551C}, no simplifications nor approximations are imposed on Eq.~\eref{eq:auto2}, and we perform the full computation of the \com{six integrals} to evaluate each $\mathbf{C}_{ij}^{\; tt'}$ element.

\subsection{Assumptions: scan strategy, focal plane layout}
\label{ssec:assumptions_ss_focal_plane}

For illustration purposes, we set a scanning strategy of the telescope, $\mathbf{\hat{r}_{s}}(t)$, to be constant in elevation ($\theta_{s}(t) = \theta_{s, 0}$) and sweeping in azimuth. As a rough approximation of usual scanning strategies adopted by CMB ground-based experiments, $\mathbf{\hat{r}_{s}}(t)$ is assumed in this section to follow a simple cosine function,
\begin{eqnarray}
	\centering
		\phi_{s}(t) = \phi_{s, 0} + \frac{\Delta az}{2}\cos\left( 2\pi f_{\rm scan} t + \psi\right),
	\label{eq:scan_strat_def}
\end{eqnarray}
where we set $\Delta az$ to be the angular size of the scan (usually of the order of a couple of degrees) and $f_{{\rm scan}}$ the scan frequency, defined as
\begin{eqnarray}
	\centering
		f_{{\rm scan}}\,[{\rm Hz}]\,\equiv\,\frac{ ss\,[\deg\,s^{-1}]}{\Delta az\,[\deg]},
	\label{eq:fscan_def}
\end{eqnarray}
where $ss= |\mathbf{ss} |$ is the norm of the scanning speed.
For simplicity, we only consider a single detector in this section, pointing along the central line of sight $\mathbf{\hat{r}_{s}}(t)$ of the telescope. Throughout this paper, this detector will be denoted $i=j=0$. Note that in section~\ref{sec:atmosphere_estimation}, in which the modeled atmospheric contamination is compared to real data, $\mathbf{C}_{ij}^{\; tt'}$ is computed using the real pointing of the experiment $\left\{ \phi=az(t),\theta=el(t)\right\}$.

\subsection{Implementation}

Significant computational resources are required to compute Eq.~\eref{eq:auto2}.
We use a Quasi Monte Carlo integration method to numerically estimate the six-integral included in the computation of a single $\mathbf{C}_{ij}^{\; tt'}$ element.  The number of quasi-random samples has been optimized and turned out to be $\sim 10^2$ for the computation of a single $\mathbf{C}_{ij}^{\; tt'}$ element. \com{More samplings lead to sub-percent change to the integral value}. In addition, we use the spatial locality of the functions involved in the integrand to speed-up the computation, namely $B(\mathbf{\hat{r}_s}(t),\mathbf{r})$, Eq.~\eref{eq:beff}, and $\chi_1(\mathbf{\Delta r})$, Eq.~\eref{eq:chinos}: we reduce the integration bounds in Eq.~\eref{eq:auto2} to scales within the beam size or the turbulent typical length. \com{For example, we reduce $d\mathbf{r'}$ to a 3d volume of radius 
\begin{eqnarray}
	\centering
		\mathbf{\hat{r}_s}(t)\cdot\mathbf{r}\, \pm\, L_o,
	\label{eq:1sigma_volume_on_r}
\end{eqnarray}}
which is effectively the 1$\sigma$ region of the correlations induced by the turbulences. In addition, the $\left( \theta, \phi \right)$ and  $\left( \theta', \phi' \right)$ variables are integrated around 
\begin{eqnarray}
	\centering
		\left( \theta_s \pm {\rm max}\left( \theta_b, \frac{L_o}{r}\right),  \phi_s \pm  {\rm max}\left( \theta_b, \frac{L_o}{r}\right)\right).
	\label{eq:1sigma_volume_on_theta_phi}
\end{eqnarray}
More distant locations are suppressed either by the beam term $B(\mathbf{\hat{r}_s}(t),\mathbf{r})$, or by the turbulence term, $\chi_1$. \com{Outside the `'1$\sigma$'' volume defined by Eqs.~\eref{eq:1sigma_volume_on_r} and~\eref{eq:1sigma_volume_on_theta_phi}, integrand contributes to less than $5\%$ of the total integral in $\mathbf{C}_{ij}^{\; tt'}$.}
Finally, the implementation takes advantage of symmetry relations such as $\mathbf{C}_{ij}^{\; tt'} = \mathbf{C}_{ji}^{\; tt'}$.

For a given set of physical parameters $\left\{ L_o, T_0, \hdots \right\}$, the implementation of the algorithm used for this article performs the computation of a single element $\mathbf{C}_{ij}^{\; tt'}$ in $\sim 1$ ms on one CPU. A basic parallelization of the algorithm is easy to implement and efficient to compute the entire set of $\mathbf{C}_{ij}^{\; tt'}$ elements.

\subsection{Illustration of the modeling for a detector auto-correlation}
\label{ssec:results_numerical_integration}

\begin{figure}
	\centering
		\includegraphics[width=6cm]{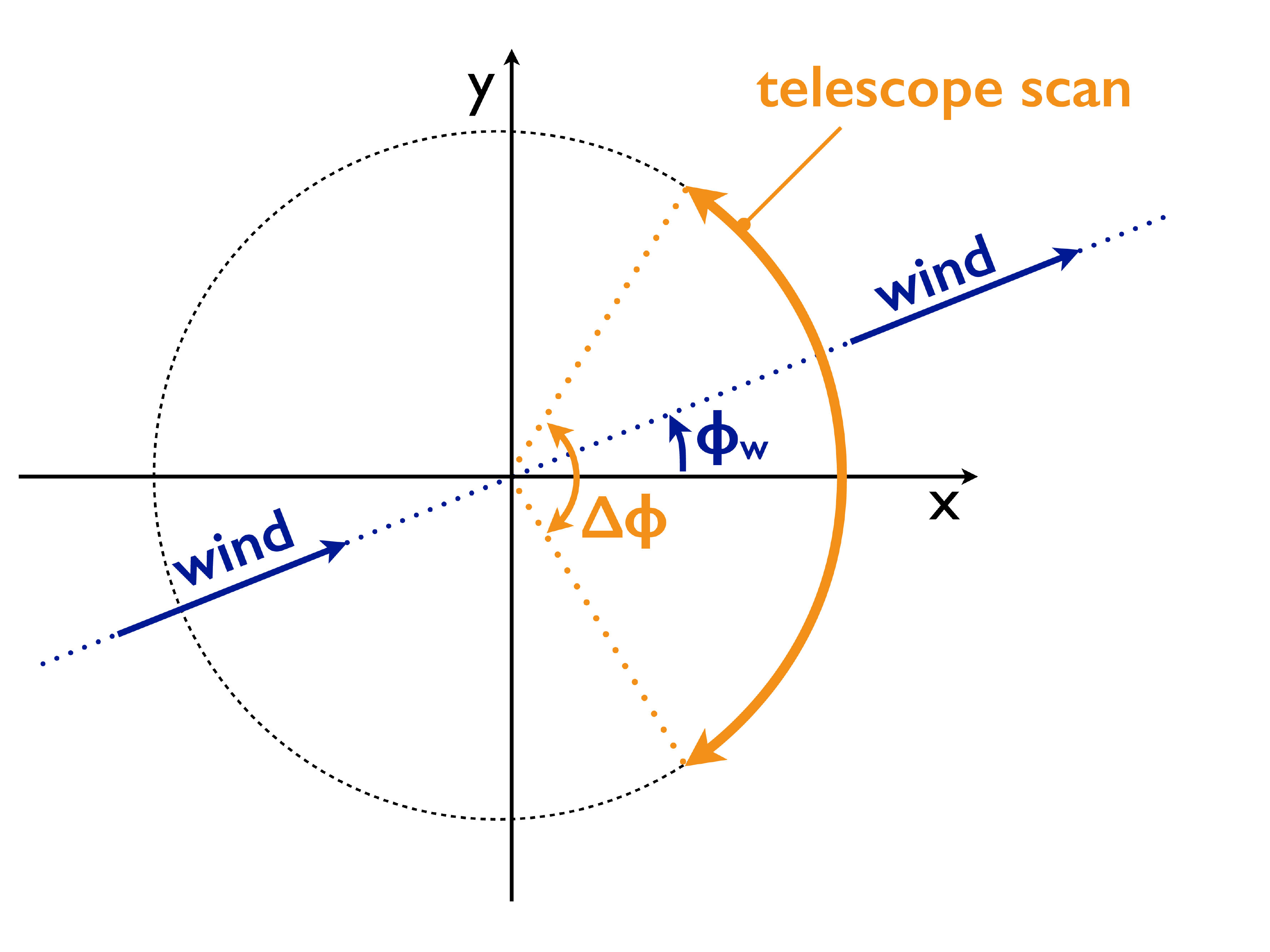}
	\caption{ Scan strategy of the telescope is a cosine function, Eq.~\eref{eq:scan_strat_def}, centered on $\phi_{s, 0} = 0$. The wind is assumed to be parallel to the $x-y$ plane and have a fiducial direction $\phi_W = 0$. }
	\label{fig:geometry_ss_wind}
\end{figure}

Assuming the specific observing  strategy given in Eq.~\eref{eq:scan_strat_def}, Fig.~\ref{fig:Loxssxt} depicts $\mathbf{C}_{ij}^{\; t t'}$ for the central detector $i=j=0$ as a function of both $t'$ ($x$-axis assuming $t=0$) and one of the parameters involved in the atmosphere conditions description ($y$-axis, all the other parameters are set to their fiducial values). $\mathbf{C}_{00}^{\; 0 t'}$ corresponds to the time-domain autocorrelation of the central detector. 
More precisely, Fig.~\ref{fig:Loxssxt} depicts the quantity $\mathbf{C}_{ij}^{\; 0  t'}/\mathbf{C}_{ij}^{\; 0 0, \, fid}$ where $\mathbf{C}_{ij}^{\; 0 0, \, fid}$ is computed for the chosen fiducial parameters $\{$ $L_o=300\, {\rm m}$, $z_0 = 2000\, {\rm m}$, $z_{\rm atm} = 40,000\, {\rm m}$, $FWHM = 3.5\, {\rm arcmin}$, $ss=1 \, \deg\,s^{-1}$, $\Delta az=5\,\deg$, $\phi_{s\,0} = 0\,\deg$, $\theta_{s,\,0} = 45\,\deg$, $\psi= 0\,\deg$ $\}$ and $\{$ $W = 25$ m s$^{-1}$,  $\phi_{W} = 0\,\deg$ $\}$ in the windy cases (panels (5) and (6) only).
A common feature across all panels in Fig.~\ref{fig:Loxssxt} is the presence of nearly scan-synchronous signals, induced by atmospheric structures, roughly acting as a sky-like signal for low wind speed and/or short period of time. We describe below the effect of changing a parameter individually on the auto-correlation function: 

\begin{itemize}
\item \textbf{Turbulence typical length, $L_o$}: panel (1) shows that an increase of this length results in a larger width of the scan-synchronous features. In the limit of an infinitely large turbulent typical scale, all the detectors would be 100\% correlated, for an infinitely long time if there is no wind. Contrarily, a $L_o \rightarrow 0$ m reduces the nearly scan-synchronous features, bringing the averaged correlation to negligible levels.

\item \textbf{Correlation amplitude, $T_0$}: as mentioned earlier, this is an effective ground temperature and its square is the global normalization of the correlation matrix $\mathbf{C}_{ij}^{\; tt'}$, Eq.~\eref{eq:auto2}, in the case of monochromatic detectors. Therefore $T_0$ simply scales the entire correlation function, and is not illustrated in Fig.~\ref{fig:Loxssxt}.

\item \textbf{Telescope scanning properties, $ss$ and $\Delta az$}: they impact $f_{\rm scan}$, Eq.~\eref{eq:fscan_def}, and consequently the frequency of the scan synchronous features in $\mathbf{C}_{ij}^{\; tt'}$. As shown in panel (2), in the absence of wind, the larger the scan speed is, the more scan-synchronous features are present. As depicted in panel (3),  the larger $\Delta az$ is (for a fixed scan speed), the fewer scan-synchronous features appear. A long and slow (respectively short and rapid) movement of the telescope modulates the atmospheric structures to low (respectively high) time streams frequencies.  

\item \textbf{Typical altitudes $z_{\rm atm}$ and $z_0$}: they have similar impact on the correlation. Panel (4) of Fig.~\ref{fig:Loxssxt} shows the effect of changing $z_0$ on $\mathbf{C}_{00}^{\; 0t'}$: smaller water vapor column results in smaller correlation amplitude, with a mild impact due to the form of $\chi_2$, cf. Eq.~\eref{eq:chis}. In addition, note that the position of the water vapor densest region would also play a role on the width of the nearly scan-synchronous features. The $\chi_2$ term acts as a weight for the contribution of turbulent structures to the atmospheric signal: the lower is $z_0$, the more important is the contribution of atmospheric fluctuations at lower altitudes.

\item \textbf{Wind properties}:  results are depicted on panels (5) and (6). The first one shows the effect of wind speed on $\mathbf{C}_{00}^{\; 0t'}$, for a constant wind direction $\phi_W = 0\,\deg$, cf. the geometry illustrated in Fig.~\ref{fig:geometry_ss_wind}. The larger is the wind speed, the more rapidly harmonics of  $\mathbf{C}_{00}^{\; 0t'}$ are suppressed. Since wind is displacing atmospheric turbulences, any detector cannot observe the same sky features at a given azimuth pointing. From Eq.~\eref{eq:auto2}, having a non-zero $W$ leads to $\mathbf{C}_{ij}^{\; tt'} \propto e^{-(W|t'-t|)^2 /2L_o^2}$: forgetting about the cross-term in this expression, this is why the correlation is suppressed for long time intervals $|t'-t|\gg L_o/W$.

Panel (6) shows the effect of a wind direction $\phi_W$ change at a given wind speed, $W\,=\,25\, m\,s^{-1}$. The geometry of the scanning strategy with the wind direction is depicted in Fig.~\ref{fig:geometry_ss_wind}: as defined in Eq.~\eref{eq:scan_strat_def}, the assumed scanning strategy is centered around $\langle \phi (t) \rangle_t = 0\,\deg$ since we took the fiducial $\phi_{s,0} = 0\, \deg$. 
On one hand, pseudo-harmonic structures of $\mathbf{C}_{00}^{\; 0 t'}$ are efficiently suppressed for $\phi_W\,\in\,\left\{ -180,\, 0,\, +180 \right\}\, \deg$. For the chosen scanning amplitude $\Delta az = 5\,\deg$, these wind directions are roughly orthogonal to the scanning direction, which means that the wind efficiently brings new, and hence uncorrelated, atmospheric fluctuations across the telescope line of sight.
On the other hand, even if atmospheric structures are shifted and suppressed as a function of $\Delta t$, significant features remain in the time streams for the case $\phi_W = \pm 90 \deg$. The physical interpretation of this effect is that the telescope resamples already scanned structures on the sky: the $\phi_W = \pm 90 \deg$ wind shifts atmosphere turbulences in a nearly parallel direction to the telescope scans. Specifically, given our assumptions for the scanning strategy, a $\phi_W = - 90 \deg$ would move atmospheric structures with the line of sight motion during the $1$st, $3$rd, ..., $(2i^{th}+1)$ subscan of the telescope. On the contrary, a $\phi_W = + 90 \deg$ would move fluctuations with the line of sight motion during the $2$nd, $4$th, ..., $2i^{\rm th}$ subscan. \cite{2005ApJ...622.1343B} gives a detailed description of this phenomenon.
\end{itemize}
The remarks above hold for cross-correlations, i.e., for $i\neq j$ case. However, we always have $\mathbf{C}_{i\neq j}^{\; t=t'} \leq \mathbf{C}_{i=j}^{\; t=t'}$ and the wind can impact correlations amplitude and shape between detectors, depending on their specific positions on the focal plane, as well as on the scanning strategy.

\begin{figure}
	\centering
		\includegraphics[width=\columnwidth]{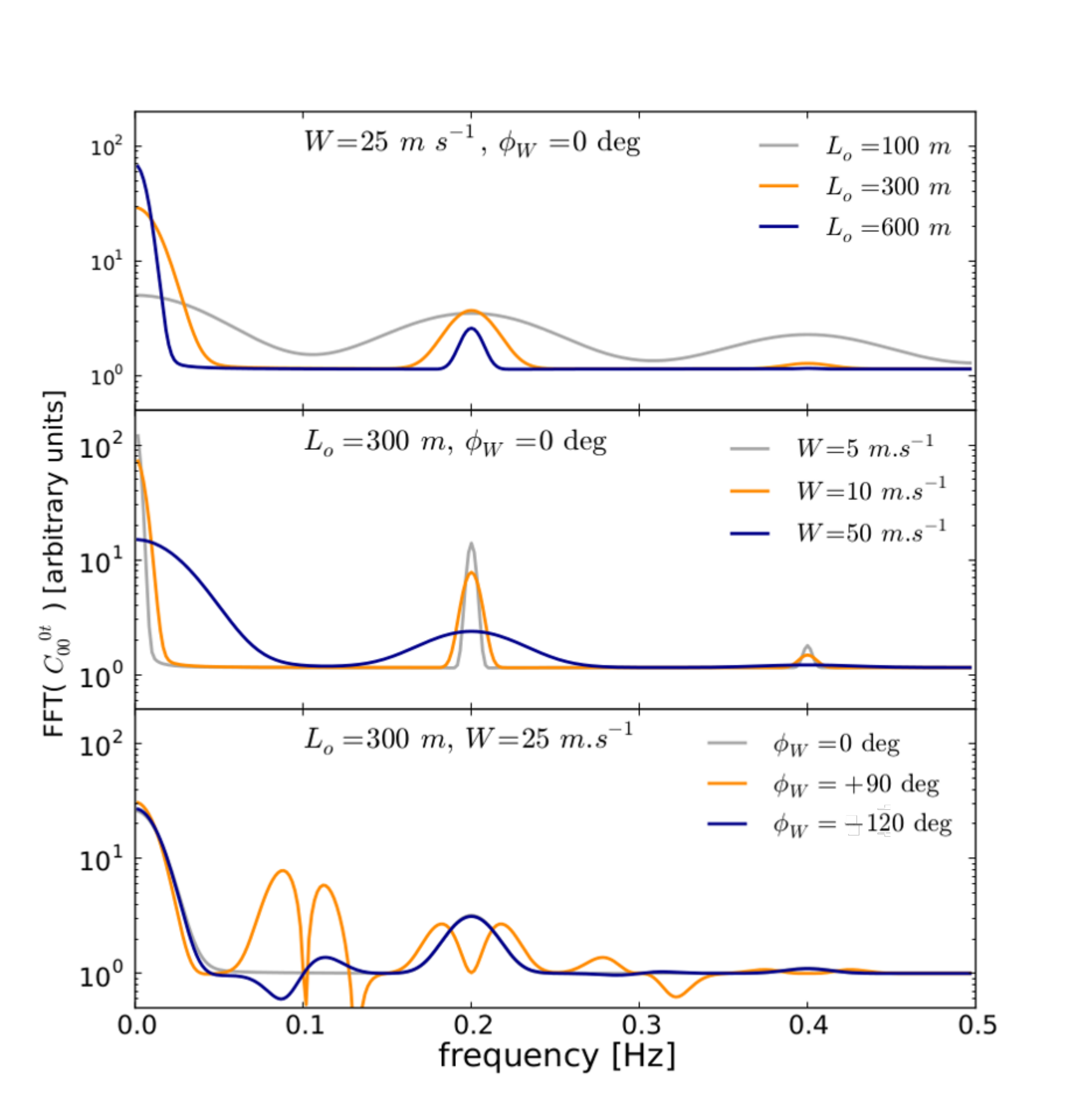}
		\caption{Auto power spectra, obtained as the Fourier transform of $C_{00}^{\;0t'}$, shown here for various $L_o$ (upper panel, assuming $W=25$ m$^{-1}$ and $\phi_W=0\,\deg$), various wind speeds (middle panel, assuming $\phi_W=0\,\deg$ and $L_o=300$ m) and various wind directions (lower panel, assuming $W=25$ m s$^{-1}$  and $L_o=300$ m).}
	\label{fig:Cijf_vs_wind}
\end{figure}

Complementary to Fig.~\ref{fig:Loxssxt} which illustrates the correlation \com{in the time domain}, Fig.~\ref{fig:Cijf_vs_wind} shows the Fourier transform of several $C_{00}^{\;0t'}$, i.e., the auto-spectra, for various wind speed and wind directions.  
First, the upper panel of Fig.~\ref{fig:Cijf_vs_wind} shows the effect of various turbulence scales $L_o$ at a fixed wind direction and speed ($W=25$ m s$^{-1}$ and $\phi_W = 0\,\deg$): similarly to the observations derived from $C_{00}^{\;0t'}$, the auto spectra present features appearing at $f_{\rm scan}=0.2$ Hz. Large $L_o$ \com{results in more power at low frequencies while} small $L_o$ leads to a broader spectrum, with spread power at high frequencies.
Second, the middle panel shows the effect of various wind speeds at a fixed wind direction and $L_o$ ($\phi_W = 0\,\deg$ and $L_o=300$ m). The larger the wind speed, the broader are the peaks in frequency space. As mentioned previously, wind displaces atmospheric structures and the detector does not resample these \com{structures at a constant frequency $f_{\rm scan}$.}
Third, the lower panel shows the effect of various wind directions at a fixed wind speed and $L_o$: as suggested by the correlation behavior in time domain, wind direction modulates the typical frequencies of atmospheric contamination features.\\

To summarize, we presented the \com{dependence} of the autocorrelation \com{regarding} the turbulence typical size, on the observing strategy, on the wind properties and on the water vapor column typical height. In the limit of a low wind speed, the model predicts the presence of nearly scan synchronous features in the detector time streams. However, the width of these features strongly depends on the typical size $L_o$ of \com{the turbulence}, and depends weakly on the typical altitudes $z_0$ and $z_{\rm atm}$. Moreover, the amplitude of these features usually decreases rapidly as a function of the elapsed time $|t'-t|$, mostly driven by the wind speed. The pseudo-periodicity of these features is modulated by the wind direction relative to the scanning strategy.\\

The next section will quantitatively compare the predictions of the modeling described above with recent observations of the CMB ground-based experiment \textsc{polarbear-i}.

\section{Quantitative comparison between the modeling predictions and real CMB data sets}
\label{sec:atmosphere_estimation}

In this section, we fit real \textsc{Polarbear-i} data sets with the modeling introduced in section~\ref{sec:atmosphere_physical_model} and derive distributions of atmosphere parameters such as typical turbulence scale and wind speed. Our approach here is quite similar to \cite{2005ApJ...622.1343B}, but differs in two aspects: first, we use the 3d modeling of the atmosphere and second, we estimate the physical atmospheric parameters through the maximization of a parametric likelihood based on the correlation matrices $\mathbf{C}_{ij}^{\,tt'}$. A quantitative comparison with previous results derived from the 2d modeling approach above the Atacama desert (\cite{2000ApJ...543..787L}) is also presented.
This analysis involves an estimator of the full covariance of the real data sets introduced in section~\ref{ssec:dijt_from_real_data}, and a parametric maximum-likelihood framework detailed in section~\ref{ssec:formalism}. Results of the fit are summarized in section~\ref{ssec:results}.

\begin{figure} 
	\centering
		\includegraphics[width=\columnwidth]{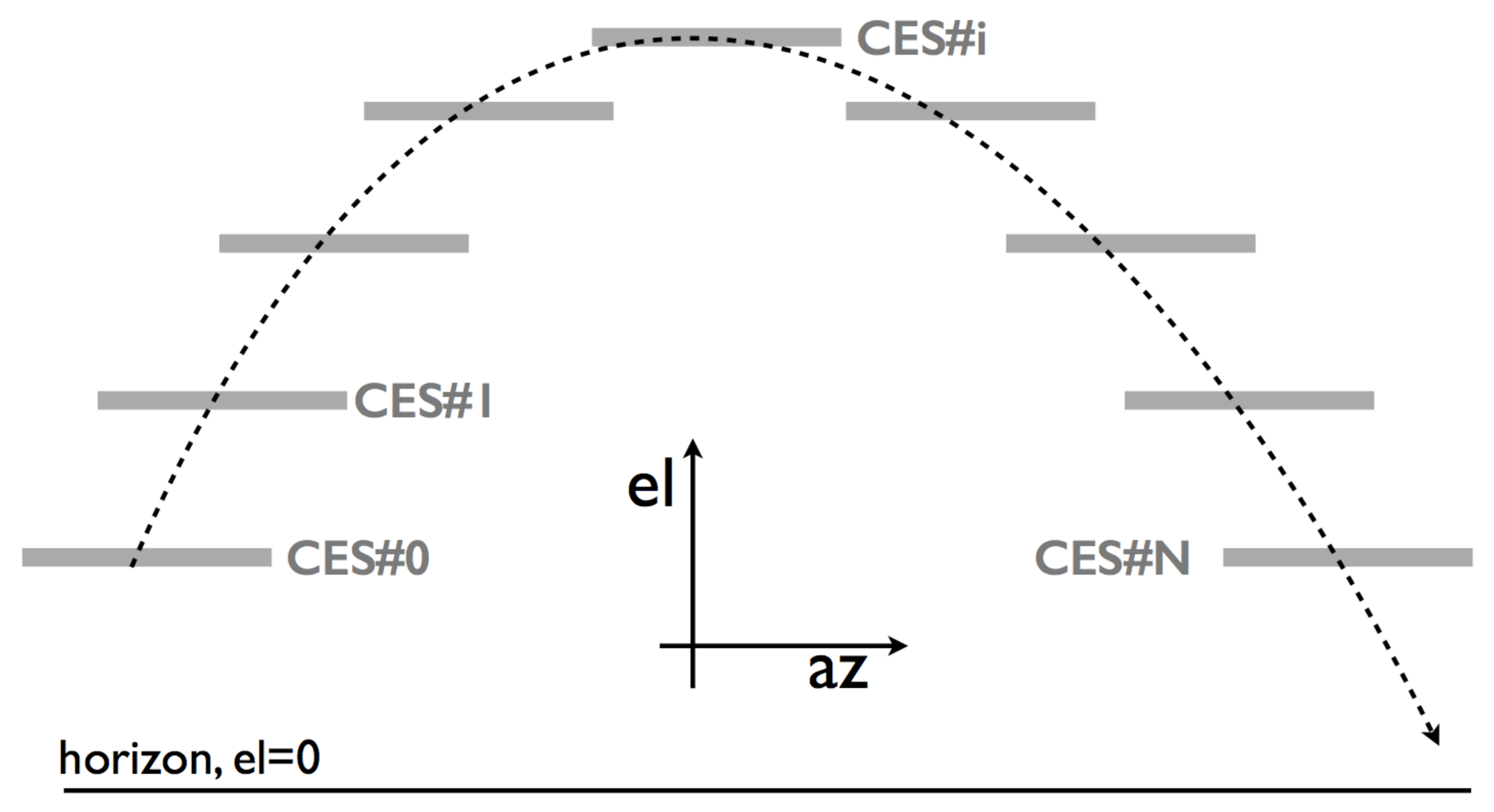}
	\caption{Drawing of the scanning strategy of $\sim\,8$-hour \textsc{polarbear-i} observations. It is composed of several constant elevation scans (CES). }
	\label{fig:PB_scanning}
\end{figure}

\subsection{Analysis of POLARBEAR-I data sets}
\label{ssec:dijt_from_real_data}

\textsc{polarbear-i} is a CMB polarization imaging experiment\footnote{http://bolo.berkeley.edu/polarbear/} located on the Chajnantor plateau, at $17,000$ feet, near to (and hence looking through a similar atmosphere as) the Atacama Cosmology Telescope\footnote{http://www.princeton.edu/act/}, the ALMA\footnote{http://www.almaobservatory.org/} and the APEX\footnote{http://www.apex-telescope.org/} observatories. It started its operations in early 2012 and has performed observations of three patches of the sky at $150$ GHz, leading to an effective coverage of $30\,\deg^2$ of the sky. The \textsc{polarbear} collaboration has recently published the results of the first season of observation (2012-2013), and demonstrated measurements of CMB $B$-modes at sub-degree angular scales (\cite{2014PhRvL.112m1302A, 2013arXiv1312.6646P, 2014arXiv1403.2369T}).\\
\textsc{polarbear-i} is composed of $1274$ bolometric detectors ($637$ dual-polarization pixels sensitive to total intensity) sampled at $190.73$ Hz (\cite{ziggy_SPIE}). As schematically depicted in Fig.~\ref{fig:PB_scanning}, \textsc{polarbear-i} CMB observations are composed of constant elevation scans (denoted CES hereafter), each of them lasting $\sim 15$ minutes. Each CES is itself composed of $\sim 100$ subscans i.e., right-to-left and left-to-right going scans, with a $\Delta \phi\sim 5\,\deg$ amplitude. The motion of the telescope is a bit more complex than the one considered in Eq.~\eref{eq:scan_strat_def}, with consecutive acceleration/deceleration (turnarounds) and constant velocity periods. \com{The scanning strategy $\left(\theta_s(t),\,\phi_s(t)\right)$ is an input to our modeling, and we use the true motion of the \textsc{Polarbear-i} telescope in this section.} The instrument periodically scans in azimuth and, as assumed in section~\ref{sec:numerical_computation}, for a given CES we are in a $\theta_s = {\rm constant}$ case. 
In this work, we fit each CES independently: each of them is considered as an independent realization of the atmospheric fluctuations, a new set of atmospheric parameters (except $z_0$ and $z_{\rm atm}$ as explained below) to be estimated and therefore a new likelihood to be maximized.\\

For a given \textsc{polarbear-i} CES, we estimate the \com{full real datasets'} covariance matrix as
\begin{eqnarray}
	\centering
		\mathbf{D}_{ij}^{\; tt'} &\equiv& \left\langle \left\langle d_i^{ (m ,\, S)} d_j^{ (n ,\, S)}\right\rangle_{\left\{ m,\,n\right\}\,\in\,  \left\{ t, \, t'\right\} } \right\rangle_{S}
	\label{eq:Dijtau_def}
\end{eqnarray}
where $d_i^{ (m ,\, S)}$ denotes the calibrated total intensity data from a pixel $i$ of the focal plane, at a given time sample $m$---sample which comes from a group of subscans $S$. The average $\langle\,\cdot\,\rangle_S$ is taken over several periods made of consecutive subscans. In each of these groups, we consider an even number of subscans, so that left-going \com{(right-going)} subscans are always averaged with left-going \com{(right-going)} subscans only.

Given the instrument specifications, the complete $\mathbf{D}_{ij}^{\; tt'}$ would have a size $( n_{\rm pixels} \times n_{\rm samples\, \in\, S} )^2 \sim (10^8)^2$. It would therefore be challenging to quickly perform operations on this data set. To simplify the computation of the real data covariance matrix $\mathbf{D}_{ij}^{\; tt'}$, we use some properties of the atmospheric contamination. For reasonable scanning strategies, since the atmosphere is evolving at rather low temporal ($\leq 1$ Hz) and spatial frequencies ($\geq 0.5\deg$), \com{a few detectors among the focal plane with low frequency} time streams can capture enough information about atmospheric fluctuations properties. \com{We hence} choose ``cardinal" pixels across the focal plane (e.g., at the center and edges), and we downsample the time streams of each of these pixels in order to lighten $\mathbf{D}_{ij}^{\; tt'}$ and therefore speed up the analysis without degrading the parameters estimation. \\

First the compression of $\left\{ m,\,n\right\}$ into temporal bins $\left\{ t,\,t'\right\}$ in Eq.~\eref{eq:Dijtau_def} corresponds to the downsampling process. Second, $d_i^{\; m}$ in Eq.~\eref{eq:Dijtau_def} is also the result of an average of the TOD over some regions of the focal plane, precisely over detectors close to the ``cardinal" pixels i.e.,
\begin{eqnarray}
	\centering
		d_i^{\; m}\rightarrow \left\langle d_k^{\; m} \right\rangle_{k\,\in\,{\rm neighbor\ pixels\ to\ pixel\ i}},
	\label{eq:neighbor_average}
\end{eqnarray}
where the neighbor pixels $k$ are the nearest observing detectors to a ``cardinal" pixel $i$.
Complementary to Eq.~\eref{eq:Dijtau_def}, the error bar on each $\mathbf{D}_{ij}^{\; tt'}$, denoted $\mathbf{\Delta D}_{ij}^{\; tt'}$, is taken as the standard deviation of 
\begin{eqnarray}
	\centering
		&&\left\langle d_i^{ (m ,\, S)} d_j^{ (n ,\, S)} \right\rangle_{\left\{ m,\,n \right\}\,\in\, \left\{ t, \,t'\right\} }\nonumber,\end{eqnarray}
computed over groups of subscans $S$.
We compute a single $\mathbf{D}$ matrix for each \textsc{polarbear-i} CES.\\

Atmospheric contamination encapsulated in $\mathbf{D}_{ij}^{\; tt'}$ can sometimes be approximated by a stationary process, meaning that $\mathbf{D}_{ij}^{\; tt'} \simeq \mathbf{D}_{ij}^{\; 0 \Delta t}$, with $\Delta t \equiv t'-t$. This pseudo-stationarity is exploited in Appendix~\ref{sec:approx_cov}, where an approximation of $\mathbf{D}_{ij}^{\; tt'}$ is introduced \com{for rapid} characterization of colored and correlated noise.
However, wind usually breaks stationarity since it modulates the atmospheric signal periodicity, which depends on the pointing of the telescope and on the direction of the wind. For example, in the case of a wind nearly parallel to the scanning strategy ($\phi_W = \pm90\;\deg$ in Fig.~\ref{fig:geometry_ss_wind}), the detectors of a left going focal plane or of a right going focal plane would not show the same correlation pattern, cf. section~\ref{ssec:results_numerical_integration}.

\begin{figure*}[htb!]
	\centering
		\includegraphics[width=15cm]{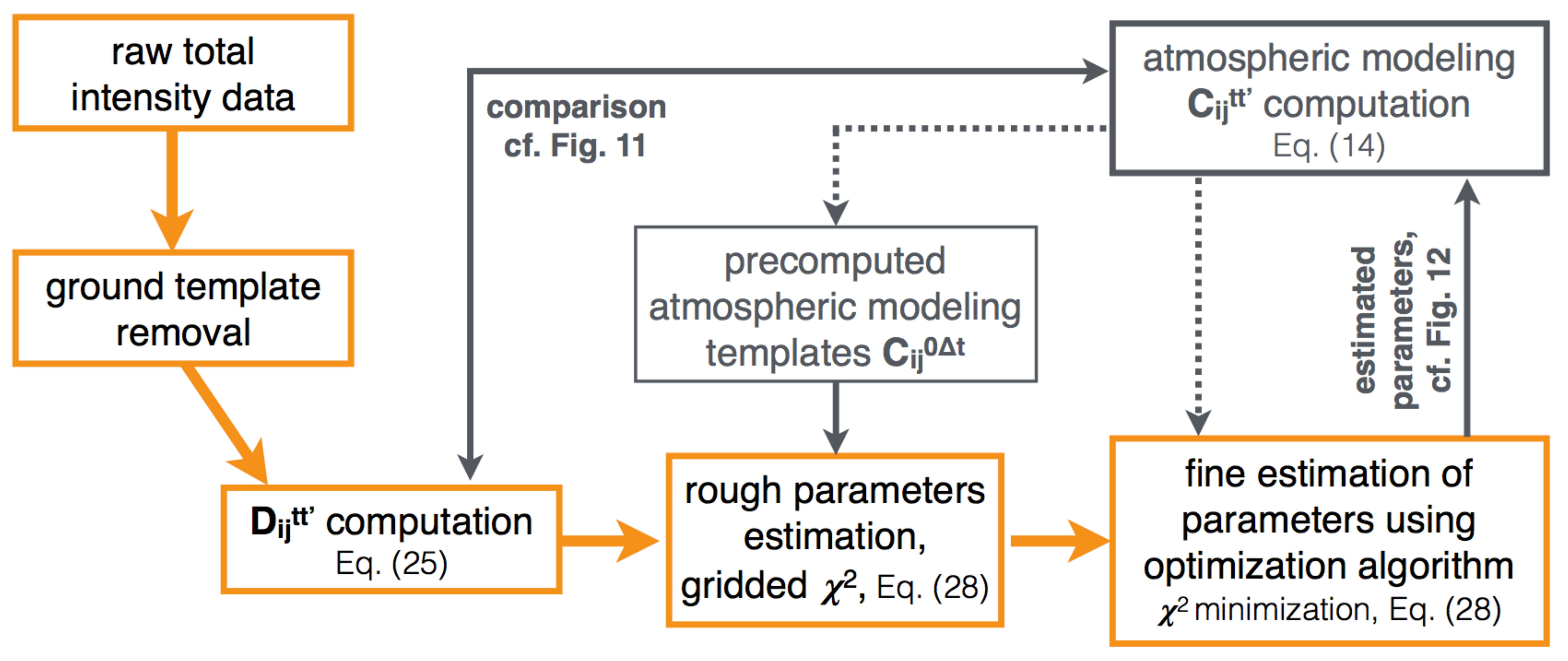}
	\caption{Schematic illustration of the algorithm we used to estimate the parameters of the atmospheric modeling from the raw data. We first filter a ground template from the raw total intensity time streams, and then compute the correlation $\mathbf{D}_{ij}^{\ tt'}$ following Eq.~\eref{eq:Dijtau_def}. A second step consists in comparing $\mathbf{D}_{ij}^{\;0\Delta t}$ with precomputed $\mathbf{C}_{ij}^{\;0\Delta t}$ (lighter on disk), leading to a rough estimate of the optimal parameters. This is the starting point for a minimization of the likelihood given in Eq.~\eref{eq:logL_parameter_estimation}, over constrained variables $p=\{$$L_o$, wind speed, wind direction, $T_0$$\}$, using gradient information in a truncated Newton algorithm, c.f.~\cite{1984SJNA...21..770N}, and based on the full matrices $\mathbf{C}_{ij}^{\;tt'}$ and $\mathbf{D}_{ij}^{\;tt'}$. Finally, we compute the final $\mathbf{C}_{ij}^{\;tt'}$ from the estimated atmospheric parameters. As an illustration, comparison of several $\mathbf{D}_{ij}^{\;0\Delta t}$ and their adjusted $\mathbf{C}_{ij}^{\;0\Delta t}$ are shown in Fig.~\ref{fig:illustration_cijt_dijt_nss4_20120923_025302}.}
	\label{fig:schematic_parameters_estimation}
\end{figure*}

\subsection{Parametric likelihood and implementation}
\label{ssec:formalism}

We want to construct a likelihood to quantitatively and robustly compare the predictions of the atmospheric contamination modeling, $\mathbf{C}_{ij}^{\;tt'}=\mathbf{C}_{ij}^{\;tt'}(p)$, function of the physical parameters $p=\{$$L_o$, $T_0$, $\mathbf{W}$,...$\}$ describing the atmosphere conditions, with the real-data matrices,  $\mathbf{D}_{ij}^{\; tt'}$, estimated from the \textsc{polarbear-i} first season data sets.\\

\com{The comparison of modeled and real covariances has been tackled in other astrophysical contexts, for example,} in the case of astrophysical foregrounds separation from multi-frequency data sets. Following the formalism by \cite{2001ITSP...49.1837P}, the negative log-likelihood $\mathcal{L}$ can be written as
\begin{widetext}
\begin{eqnarray}
	\centering
		-2\log\left( \mathcal{L}( p ) \right) &\propto& \sum_{t,t'}{ \left\{  {\rm tr}\left( \mathbf{C}_{ij}^{\;tt'}(p)\left(\mathbf{D}_{ij}^{\;tt'}\right)^{-1} \right) - \log\left[  \det\left(  \mathbf{C}_{ij}^{\;tt'}(p)\left(\mathbf{D}_{ij}^{\;tt'}\right)^{-1} \right) \right]  - {\rm constant }\right\}}\label{eq:logL_parameter_estimation1}\\
		&\simeq&  \sum_{t,t'}{ \left\{  {\rm tr} \left[ \left(\mathbf{C}_{ij}^{\;tt'}(p) - \mathbf{D}_{\;ij}^{\; tt'}\right) \left(\mathbf{D}_{ij}^{\;tt'}\right)^{-1} \left(\mathbf{C}_{ij}^{\;tt'}(p) - \mathbf{D}_{ij}^{\;tt'}\right) \left(\mathbf{D}_{ij}^{\;tt'}\right)^{-1} \right]\right\} }, 
	\label{eq:logL_parameter_estimation}
\end{eqnarray}
\end{widetext}
where Eq.~\eref{eq:logL_parameter_estimation} is the quadratic approximation of Eq.~\eref{eq:logL_parameter_estimation1} (\cite{2003MNRAS.346.1089D}).
For the computation of  $\mathcal{L}(p)$, note that $\mathbf{C}_{ij}^{\; tt'}( p )$ is given by the integration of Eq.~\eref{eq:auto2} for a given set of parameters $p$, cf. section~\ref{sec:numerical_computation}, and $\mathbf{D}_{ij}^{\; tt'}$ is the covariance matrix computed from \textsc{polarbear-i} real data sets, cf. section~\ref{ssec:dijt_from_real_data}.

\begin{figure*}
	\centering
		\includegraphics[width=8cm]{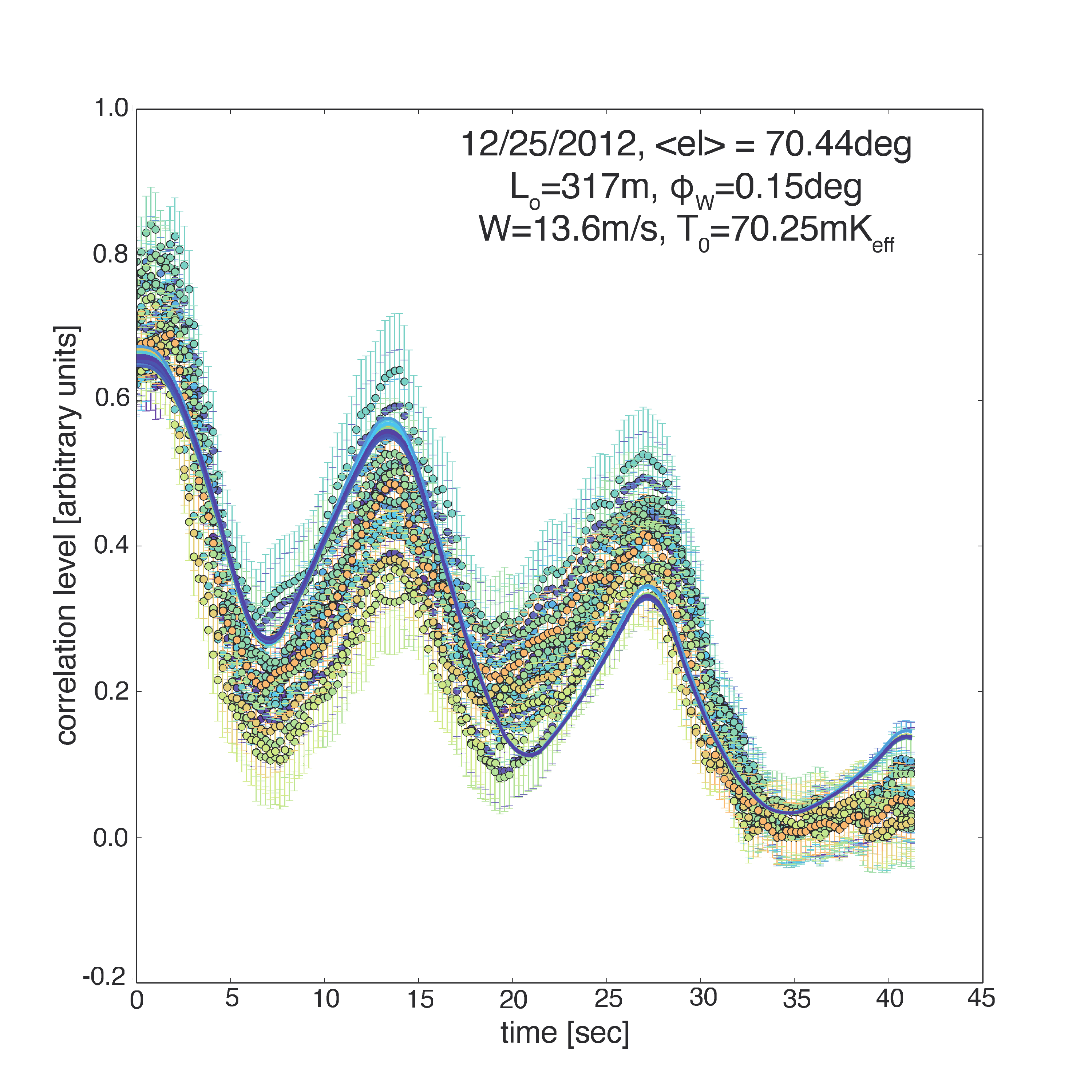}~\includegraphics[width=8cm]{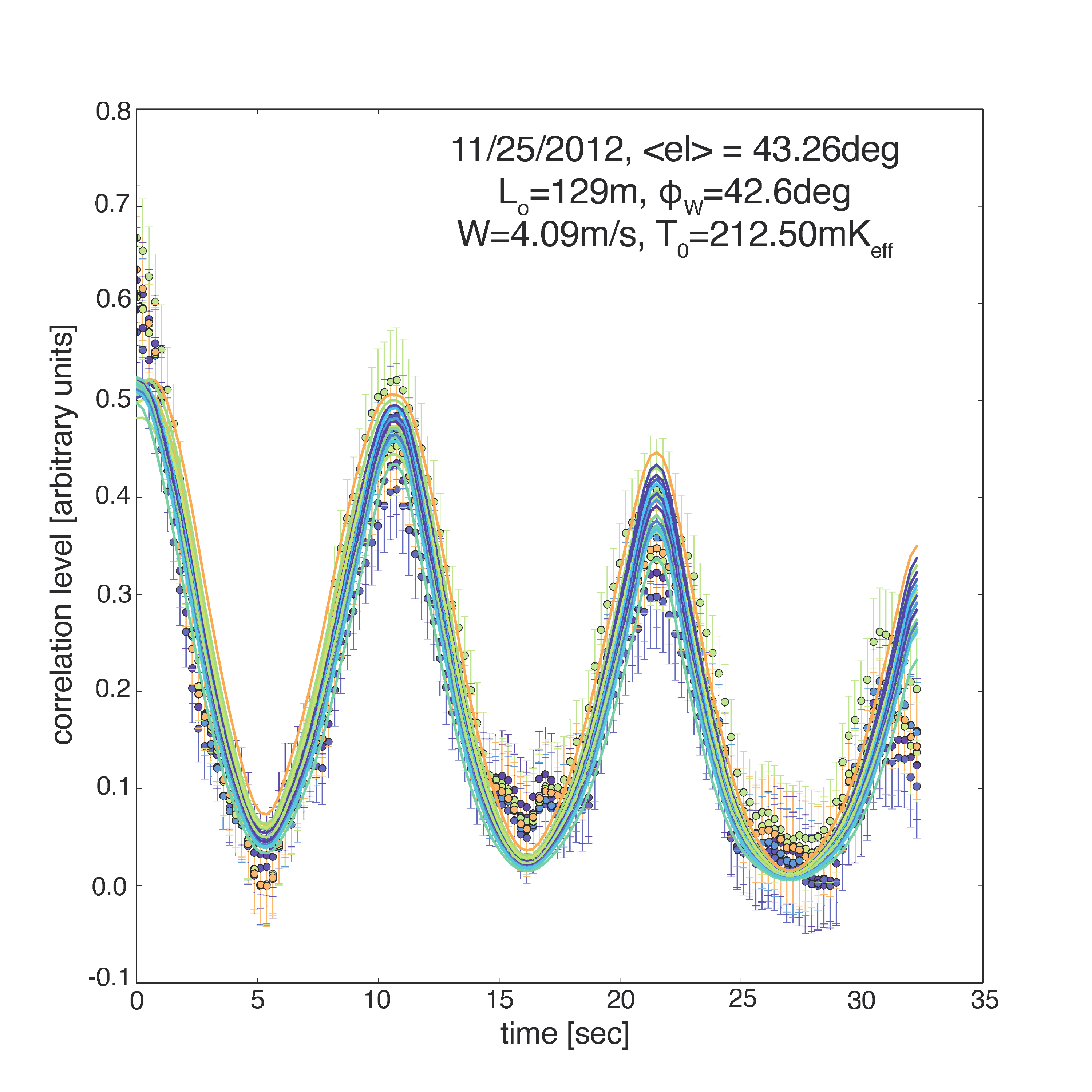}
		\includegraphics[width=8cm]{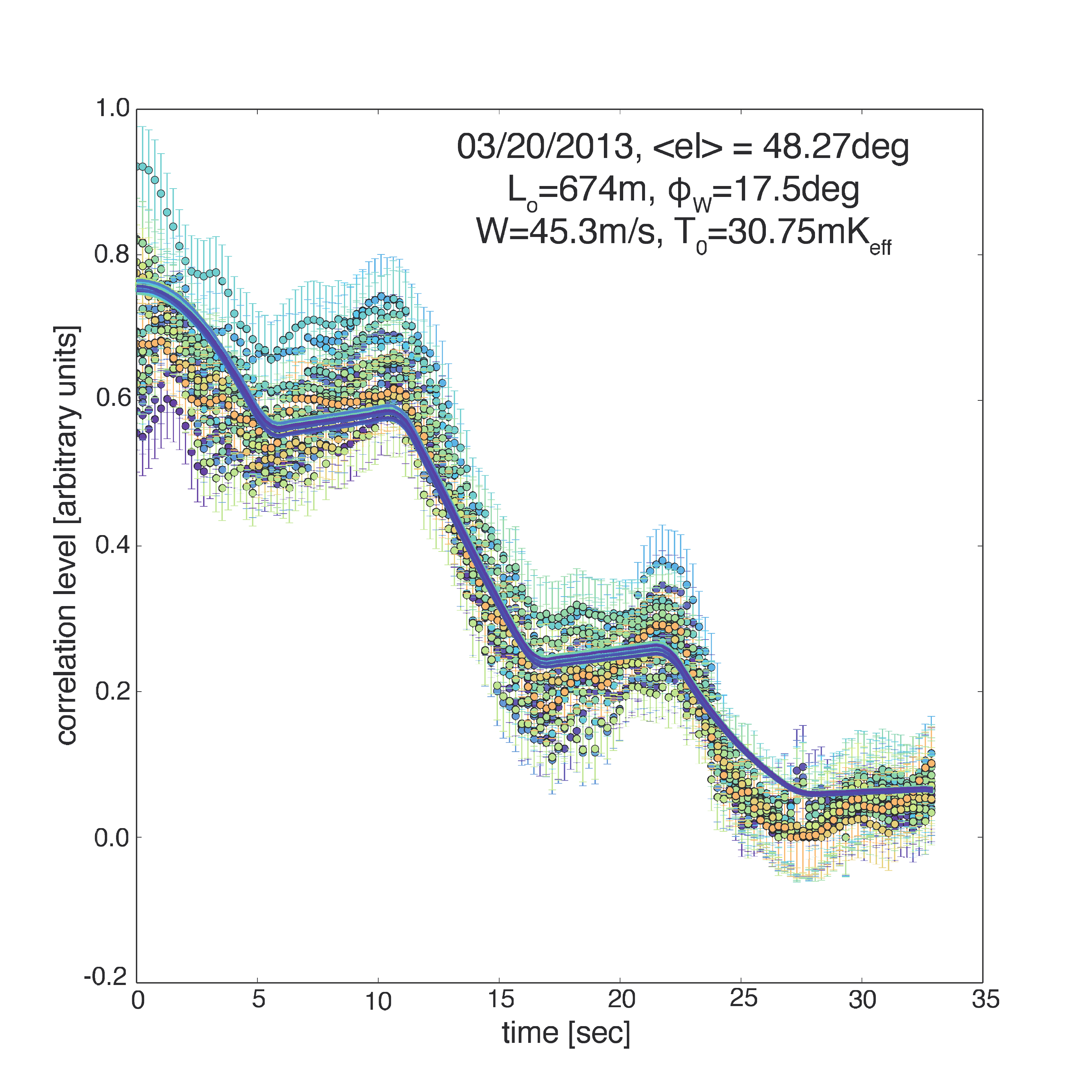}~\includegraphics[width=8cm]{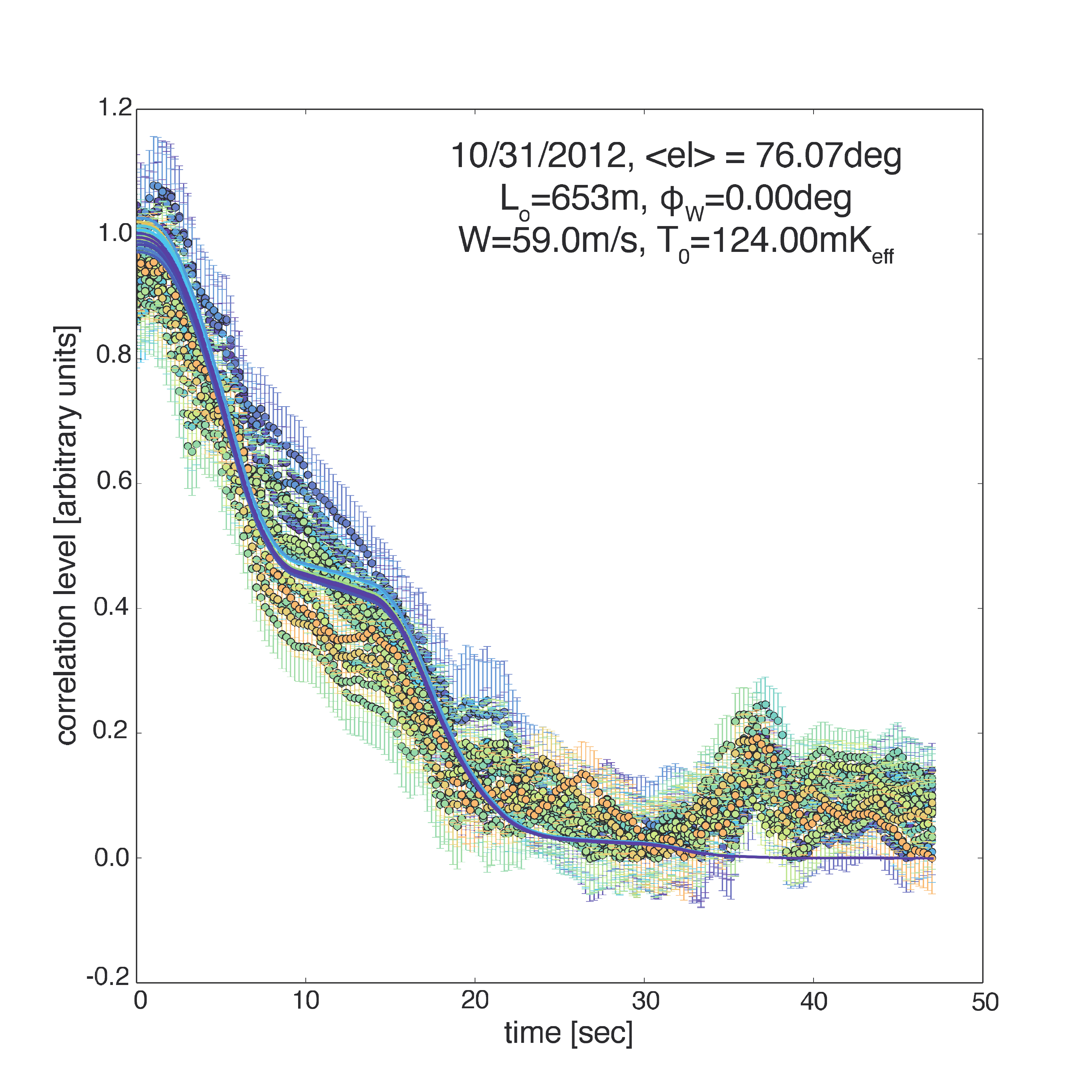}
		\caption{Example of estimated $\mathbf{D}_{ij}^{\; 0\Delta t}$ elements (colored points with error bars given by $\mathbf{\Delta D}_{ij}^{\; 0\Delta t}$), defined in Eq.~\eref{eq:Dijtau_def}. They are shown here as a function of $\Delta t = |t' - t|$, each color corresponding to a different pixels pair $\{i,\, j \}$. Solid lines are the best fit model $\mathbf{C}_{ij}^{\; 0\Delta t}$, obtained by optimizing Eq.~\eref{eq:logL_parameter_estimation} over four atmospheric parameters: the turbulence typical length ($L_o$), the wind speed and direction ($W$ and $\phi_W$) and the effective ground temperature ($T_0$). These solid lines tend to overlap in some cases, but are always in good agreement with the observational data sets. Each panel corresponds to an independent CES, and the best fit parameters are detailed in the top-right corner of each figure, together with the date of the observation and the average elevation.}
	\label{fig:illustration_cijt_dijt_nss4_20120923_025302}
\end{figure*}

For the estimation of $\mathbf{D}$, we consider time streams $d_i^{\ m}$ over groups of 6 consecutive subscans, defining $S$ in Eqs.~\eref{eq:Dijtau_def} and~\eref{eq:neighbor_average}: averages $\langle\,\cdot\,\rangle_S$ are then taken over all the available groups of 6 consecutive subscans within a CES, corresponding to $\sim 35$ s long time streams. As mentioned earlier, note that $S$ is composed of an even number of subscans, so that left-going scans are only averaged with left-going scans, cf. \cite{2005ApJ...622.1343B}. Similarly, right-going scans are only averaged over right-going scans. 
Prior to Eqs.~\eref{eq:Dijtau_def} and~\eref{eq:neighbor_average}, we remove the global mean signal from all the TOD over the course of a given CES. A ground-template, obtained as the estimation of an azimuth-fixed signal, is removed as described by \cite{2014arXiv1403.2369T}.
In addition, we choose $5$ cardinal pixels (one at the center and four on the edge, forming a square shape). The average in Eq.~\eref{eq:neighbor_average} is taken over the 5 closest working pixels to each of the cardinal pixels. We downsample the time streams from $190.73$ Hz to $2$ Hz, \com{in order to speed up the algorithms without missing the main features of the atmospheric contamination. As shown in Fig.~\ref{fig:illustration_cijt_dijt_nss4_20120923_025302} and described in section~\ref{ssec:results}, these features are essentially pseudo scan-synchronous occurring at a frequency of $\sim 0.25$ Hz in the case of \textsc{polarbear-i}.}

In addition, we provide the optimization algorithm with the derivatives of the likelihood, with respect to the considered parameters, namely $\partial \mathcal{L}/\partial L_o$, $\partial \mathcal{L}/\partial W$, $\partial \mathcal{L}/\partial \phi_W$ and $\partial \mathcal{L}/\partial T_0$. These derivatives are semi-analytically computed using Eq.~\eref{eq:logL_parameter_estimation}, and include derivatives of the modeled covariance $\mathbf{C}_{ij}^{\;tt'}$ defined in Eq.~\eref{eq:auto2}. Having semi-analytical derivatives of the likelihood leads to a significant speed-up of the optimization. 

We implement and test the algorithms on NERSC systems\footnote{National Energy Research Scientific Computing Center machines~\textit{http://www.nersc.gov/}}, and use a simple parallelization scheme in which a single process performs the analysis of one CES; this corresponds to the computation of a $\mathbf{D}_{ij}^{\; tt'}$ and the estimation of the parameters $\{$$L_o$, $W$, $\phi_W$, $T_0$$\}$ by maximizing the likelihood given in Eq.~\eref{eq:logL_parameter_estimation}. For this work, we decide to fix $z_0=2000$ m and $z_{\rm atm}=40,000$ m as they turn out to be badly conditioned: the dependence of $\mathbf{C}_{ij}^{\; tt'}$ on these quantities is quite weak, and make the atmospheric model degenerate, in particular with the amplitude $T_0$ (section~\ref{ssec:results_numerical_integration}). We therefore allow only $\{$$L_o$, $W$, $\phi_W$, $T_0$$\}$ to vary. In addition, scan speed, elevation and azimuth involved in Eq.~\eref{eq:auto2} are set by the real observations of the telescope, i.e., we compute the modeled $\mathbf{C}$ for the detectors location used to compute $\mathbf{D}$, using the same scanning strategy, and at the same binned samples $\{ t, t'\}$.\\

Finally, as  illustrated in Fig.~\ref{fig:schematic_parameters_estimation}, the estimation of the atmospheric modeling parameters is performed in two steps: we first compare $\mathbf{D}_{ij}^{\; 0 \Delta t}$ with precomputed templates $\mathbf{C}_{ij}^{\; 0 \Delta t}$, estimated for various atmospheric conditions and saved on disk. The minimization of $-2\log (\mathcal{L} )$ gives a rough estimate of the adjusted parameters, which are used as a starting point for the second step. We maximize the likelihood expressed in Eq.~\eref{eq:logL_parameter_estimation}, based this time on the matrices $\mathbf{D}_{ij}^{\; tt' }$ and $\mathbf{C}_{ij}^{\; tt'}$, using gradient information in a truncated Newton algorithm, e.g., \cite{optimization_book}.

\subsection{Results}
\label{ssec:results}
Fig.~\ref{fig:illustration_cijt_dijt_nss4_20120923_025302} shows the correlations between the five ``cardinal" pixels,  $\mathbf{D}_{ij}^{\; 0\Delta t}\, +\, {\rm constant}\,\pm\, \mathbf{\Delta D}_{ij}^{\; 0\Delta t}$ (colored points), and the corresponding estimated predictions from the atmospheric model, $\mathbf{C}_{ij}^{\; 0\Delta t}\, +\, {\rm constant}$ (solid lines), with ${\rm constant}\,\equiv\,  - min\left(\mathbf{D}_{ij}^{\; 0\Delta t}\right)$. Only four parameters are adjusted here, namely the turbulence typical size ($L_o$), the wind speed and direction ($\mathbf{W}$) and the effective ground temperature ($T_0$). The four panels have been picked from among the analysis of the entire first season of \textsc{polarbear-i}, corresponding to $8432$ CES for the three \textsc{polarbear-i} patches (\cite{2014arXiv1403.2369T}). The depicted CES are taken on different days at different elevations (cf. title on each panel), and are chosen to show various atmospheric conditions.\\

\begin{figure*}
	\centering
	\includegraphics[width=16cm]{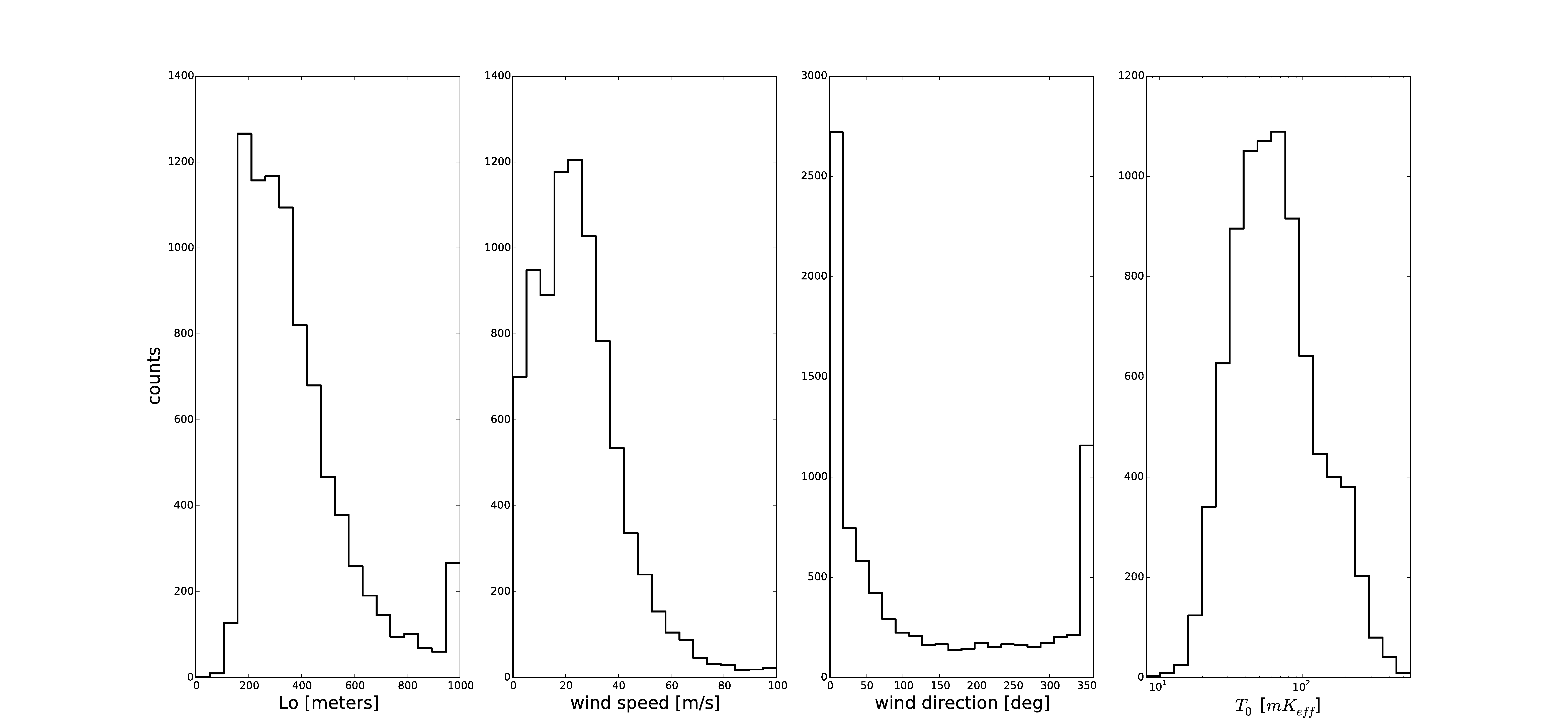}
		\caption{ Distribution of several physical parameters, recovered using the procedure described in section~\ref{sec:atmosphere_estimation}: $L_o$ (left panel), wind speed (second), wind direction (third) and effective temperature (right panel). These histograms are derived from the likelihood optimization, Eq.~\eref{eq:logL_parameter_estimation}, over each CES of the first season of \textsc{polarbear-i}.}
	\label{fig:atmosphere_detection}
\end{figure*}

As predicted by the model, nearly scan-synchronous features are present in the real data sets, with a decreasing amplitude as a function of time. 
As mentioned in section~\ref{sec:numerical_computation}, the amplitude of the curves depends on the effective temperature $T_0$, the slope of the envelope is a function of the wind speed, the width of the peaks is related to $L_o$ and their relative positions depends on the wind direction. These results show that a four-parameter physical model of the atmospheric contamination seems able to describe the full noise covariance matrix estimated from real CMB data sets.\\
A set of $\{$$L_o$, $W$, $\phi_W$, $T_0$$\}$ is estimated for each CES. Fig.~\ref{fig:atmosphere_detection} shows the distributions of each of the estimated parameters, for all the analyzed CES. For this particular \textsc{polarbear-i} data set, $L_o$, wind speed, wind direction and temperature distributions seem to have broad peaks around $\sim 300\pm100$ m, $\sim 25\pm10$ m s$^{-1}$,  $\sim 0\deg$ and $\sim 60\pm 30\,mK_{\rm eff}$ respectively. The wind direction turns out to be weakly constrained from the \textsc{polarbear-i} data sets: as mentioned in the previous section, this parameter only affects the relative position of the $\mathbf{D}_{ij}^{\;tt'}$ scan-synchronous peaks, and those can be quite small in some cases. Moreover, the small amplitude $\Delta az$ of the subscans does not offer a strong lever arm on $\phi_W$. The peaks around $\phi_W=0\,\deg$ and $\phi_W=360\,\deg$ in the wind direction histogram correspond to the lower and upper bounds imposed during the likelihood-optimization, indicating that the algorithm is unable to find better optimal coordinates given the provided data sets. \\
A comparison between $T_0$ as estimated from the CMB data set and the true ground temperature as measured by the \textsc{polarbear} weather station gives a rough estimate for the conversion factor between $K_{\rm eff}$ and $K$. An average of $0\,^\circ C$ over the year at the ground level leads to the relation $1\,mK_{\rm eff} \sim 4.55_{-1.52}^{+4.56}\,K$.\\
An interesting quantity derived above is the typical size of \com{turbulence}, $L_o$. The distribution of estimated values shows a peak around $300$ m, which is larger than the assumed $1-100$ m scale in \cite{1995MNRAS.272..551C} and lower than the observed $10$ km above the Owens Valley Radio Observatory in \cite{1997A&AS..122..535L}. However, the estimation of $L_o$ depends strongly on the site location, and measurements seem to vary significantly between different frequency range of observation. In particular, \cite{1997A&AS..122..535L} points out that interferometers on Mauna Kea or Plateau de Bure in France have observed quick phase variations, indicating the presence of rather small fluctuation scales. Realistic atmosphere simulations in the optical regime above the Atacama desert, e.g., \cite{2013MNRAS.436.1968M}, indicate that atmospheric fluctuations can take place down to the $100-500$ m scales, which is in agreement with our estimate of $L_o$.\\

\begin{figure}[htb!]
	\centering
		\includegraphics[width=\columnwidth]{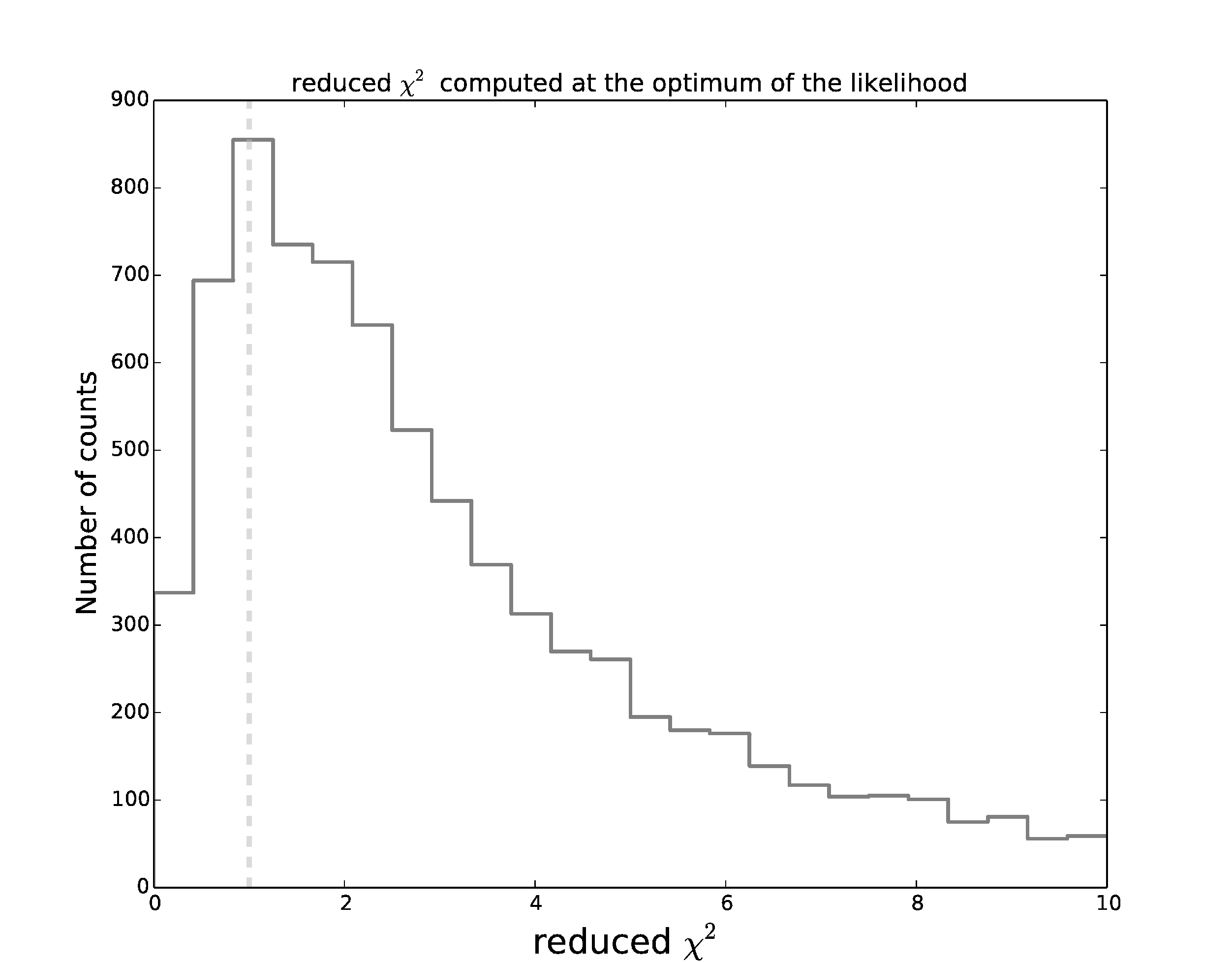}
		\caption{Distribution of reduced $\chi^2$ over all the analyzed CES, evaluated using Eq.~\eref{eq:chi2_def_after_optimization}. The dashed vertical line corresponds to a reduced $\chi^2$ of 1.}
	\label{fig:usual_chi2_distribution}
\end{figure}

To evaluate the goodness of fit of the likelihood-optimized models presented in Fig.~\ref{fig:illustration_cijt_dijt_nss4_20120923_025302}, we compute the following $\chi^2$,
\begin{eqnarray}
	\centering
		\chi^2 \equiv \sum_{i,j,t,t'}	\left(\frac{\mathbf{D}_{ij}^{\;tt'} - \mathbf{C}_{ij}^{\;tt'}(\bar{p})}{\mathbf{\Delta D}_{ij}^{\;tt'}}\right)^2,
	\label{eq:chi2_def_after_optimization}
\end{eqnarray}
where $\bar{p}$ corresponds to the likelihood-optimized atmosphere parameters. We show in Fig.~\ref{fig:usual_chi2_distribution} the distribution of the reduced $\chi^2$ computed for each CES using Eq.~\eref{eq:chi2_def_after_optimization}. This distribution peaks around $1$ as one could expect for models sensibly describing the real observations. 
We should point out that the four free parameters $\{$$L_o$, $W$, $\phi_W$, $T_0$$\}$ might be slightly degenerate between each other. Nevertheless, error bars on the estimation of these parameters, computed using the second derivative of the likelihood at the peak, Eq.~\eref{eq:logL_parameter_estimation}, give uncertainties of the order of 5-10\% of the central values.
\com{Complementary to the $\chi^2$, we also evaluated the probability to exceed (PTE). We find that 25\% (8\%) of the CES have a reduced $\chi^2$ with a PTE lower than 5\% (1\%). Fig.~\ref{fig:illustration_cijt_dijt_nss4_20120923_025302} shows cases which have a reduced $\chi^2\leq 3.0$. The significant part of the data which has a low PTE can be caused by atmospheric conditions that break our assumptions---in particular the hypothesis of frozen turbulences or the assumption that wind has a flat profile with altitude.}

\subsection{Comparison with \cite{2000ApJ...543..787L} results above the Atacama desert and with weather station measurements}
\label{ssec:comp_2d_screen}
\com{We quantitatively compare in this paragraph the predictions of our new modeling with independent measurements taken by a weather station, and with previous works which assumed the two-dimensional frozen screen approach for the atmospheric emission.}\\
\com{First,} as mentioned in appendix~\ref{sec:3d_and_2d_discussion}, we \com{can compare} the estimation for the brightness of fluctuations, $B^2_\nu\,[mK^2]$, using Eq.~\eref{eq:comparison_2d_3d} with the measurements from \cite{2000ApJ...543..787L}. This latter presents measurements from an interferometer observing the sky at $11$ GHz (converted in the article to a brightness at $40$ GHz) in the context of the 2d frozen screen approximation. This interferometer is composed of two $1.8$ m-dishes, and we will assume that the instrument observes $\gamma\,\sim1\,\deg$ angular scales on the sky. Since this latter angle and \textsc{polarbear} probed angular scales are both rather small, a comparison between $T_0$ and $B_\nu$ is possible but has to be treated carefully, cf. section~\ref{ssec:3d_2d_comparison_introduction} and appendix~\ref{sec:3d_and_2d_discussion}.
The altitude and thickness of the turbulent layer being uncertain, \cite{2000ApJ...543..787L} considers a range of possible values for $B^2_\nu$ corresponding to various values of $h$ (altitude of the turbulent layer). Even though this latter work quotes amplitudes of the fluctuations in $Ah^{-8/3}\,[mK^2$ m], we follow here the notations from \cite{2005ApJ...622.1343B}, with a brightness taken as
\begin{eqnarray}
	\centering
		B_\nu^2 \simeq Ah^{5/3}\,[mK^2],
\end{eqnarray}
which is valid in the limit of long averaging time.
Table~\ref{table:LH_vs_this_work} summarizes the values quoted in \cite{2000ApJ...543..787L}, extrapolated from $40$ GHz to $150$ GHz such as
 \begin{eqnarray}
 	\centering
		B_{150\,{\rm GHz}}^2 &=& B_{40\,{\rm GHz}}^{2}\left( \frac{\tau_{150\,{\rm GHz}}}{\tau_{40\,{\rm GHz}}}\right)^2\label{eq:scaling_40_150}\\
		&\simeq& \left( 41.1^{+22.2}_{-12.1}\right) \,B_{40\,{\rm GHz}}^{2}\ {\rm for\ PWV=1\pm1\,mm}\nonumber,
\end{eqnarray}
where $\tau$ is the opacity introduced in Eq.~\eref{eq:transmission_def}. Table~\ref{table:LH_vs_this_work} also shows the inferred values for the fluctuations brightness from \textsc{polarbear-i} data sets, using estimations of $T_0$ and Eq.~\eref{eq:comparison_2d_3d}. 
Both sets of quoted results have significant error bars, associated with the conversion from refractive index to brightness temperature in \cite{2000ApJ...543..787L}, and due to the variations of $L_o$ in our study, cf. Eq.~\eref{eq:comparison_2d_3d}. \com{In addition to providing a new semi-analytical comparison between 2d and 3d atmosphere modeling, we show that both} sets of measurements turn out to be $1\sigma$ compatible. \\ 
\bgroup
\def\arraystretch{1.75}
\begin{table*}
\centering
\begin{tabular}{c|ccc|ccc}  & \multicolumn{3}{c|}{\cite{2000ApJ...543..787L}} & \multicolumn{3}{c}{This work---$B^2_\nu$ from $T_0$, $L_o$ and Eq.~\eref{eq:comparison_2d_3d}}  \\
\hline
Quartile & $500$ m & $1000$ m & $2000$ m &  $500$ m & $1000$ m & $2000$ m \\
\hline
25\% & $123^{+128}_{-98}$ & $386^{+401}_{-307}$ & $1233^{+1286}_{-980}$ & $ 259 \pm 89$ & $ 488 \pm 179$ &  $820 \pm 357$ \\
50\% & $575^{+598}_{-457}$ & $1808^{+1881}_{-1436}$ & $5960^{+6199}_{-4735}$ & $  985 \pm 337$ & $ 1854 \pm 678$ & $ 3112 \pm 1355$\\
75\% & $2384^{+2489}_{-1893}$ & $7809^{+8123}_{-6204}$ & $24660^{+25650}_{-19590}$ & $ 3377 \pm 1155$ & $ 6358 \pm 2324$ & $ 10671\pm 4647$ \\
\end{tabular}
\caption{Brightness of the atmospheric fluctuations, in $mK^2$ above the Atacama desert, Chile. Comparison of results from \cite{2000ApJ...543..787L} (scaled from $40$ GHz to $150$ GHz using Eq.~\eref{eq:scaling_40_150}) and this work, as a function of the considered altitude for the turbulent layer $h$, and as a function of the quartile of measurements distribution. Similarly to \cite{2000ApJ...543..787L}, we assume a thickness $\Delta h = h$ for the turbulent layer. There are 50\% uncertainties on the quoted results from \cite{2000ApJ...543..787L}, associated with the conversion from refractive index to brightness temperature. We combine this error with the uncertainty associated to the conversion from $40$ to $150$ GHz due to the PWV value, Eq.~\eref{eq:scaling_40_150}. We consider a turbulent scale $L_o=300$ m for the estimation of the brightness from our estimates of $T_0$, and include $\pm\,100$ m variations around this value (cf. left panel of Fig.~\ref{fig:atmosphere_detection}) leads to $\sim\,30\%$ uncertainties on our results.}
\label{table:LH_vs_this_work}
\end{table*}
\egroup
\begin{figure*}[htb!]
	\centering
		\includegraphics[width=16cm]{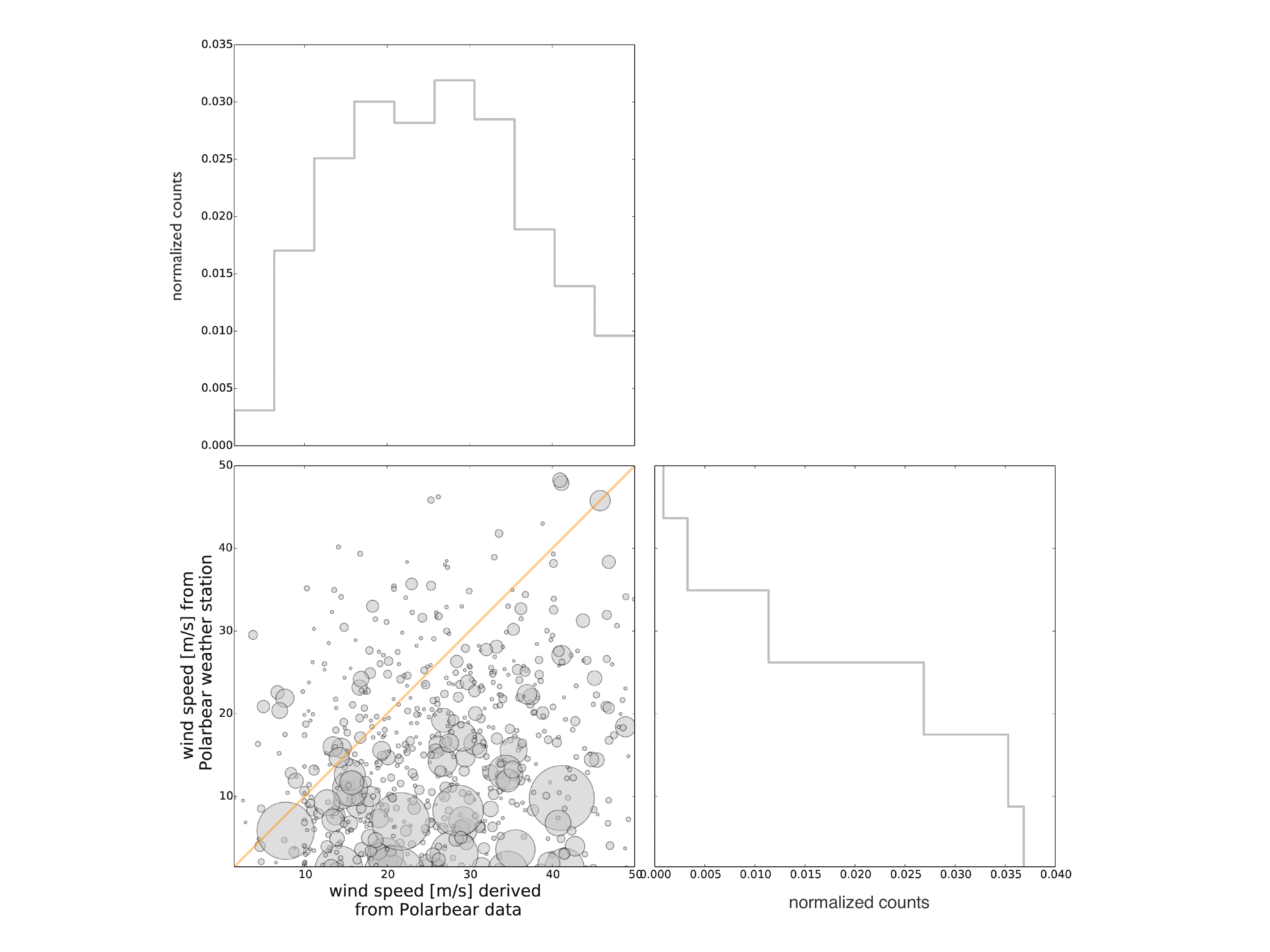}
		\caption{ Comparison of the wind speed as measured by the weather station at the ground level (vertical axis) with the atmospheric wind as estimated from the \textsc{polarbear-i} first season data sets analysis (horizontal axis). Radii of circles are proportional to the inverse of the reduced $\chi^2$, Eq.~\eref{eq:chi2_def_after_optimization}, and orange line corresponds to $x=y$. \com{The side panels correspond to the normalized distributions for the measurements taken by the weather station (right panel) and for the wind speed estimated by our analysis (top panel).} }
	\label{fig:WS_vs_PB_wv}
\end{figure*}
\com{Second}, to quantify the correlation between the atmospheric parameters estimated from the CMB data sets and the ones coming from the \textsc{polarbear} site weather station measurements, we compute the Spearman rank-order correlation coefficient, $\rho$, between the two data sets. We only use the best constrained data sets, i.e., such that their reduced $\chi^2 \leq 3.0$, Eq.~\eref{eq:chi2_def_after_optimization} and Fig.~\ref{fig:usual_chi2_distribution}. \com{In this way, we remove the worst PTE cases.} We find $\rho=46.8\%$ for the wind speed, $\rho=35.5\%$ for the wind direction and $\rho=65.2\%$ for the ground temperature.
Because the $\mathbf{C}_{ij}^{\;tt'}$ estimation gives the atmospheric properties at an altitude where the atmospheric fluctuations are located,  the correlation between the weather station and the CMB data sets is not expected to be very strong. In fact, we find \com{only a slight} correlation between the two data sets. 
To complement this comparison, we show in Fig.~\ref{fig:WS_vs_PB_wv} the wind speed values as measured by the weather station at the ground level with the wind speed estimated from the \textsc{polarbear-i} data sets analysis. Along with the distributions of each data sets, data points are shown as circles, with sizes proportional to the inverse of their reduced $\chi^2$. This illustrates similar distributions of wind speeds across the year, but with noticeable differences: (1) the wind speed as measured from $\mathbf{D}_{ij}^{\;tt'}$ is on average higher than the ground wind speed and (2) the slope of a possible linear fit would not be equal to 1. Both these remarks might be explained by the fact that wind at the ground level is not necessarily the same as the one displacing the \com{atmospheric turbulence}. This latter may take place at a different altitude, with another amplitude and potentially direction. A more relevant correlation would have to be performed between the atmospheric parameters estimated from CMB data sets and the measurements of a balloon probe above the observatory, as in \cite{2005ApJ...622.1343B}. In addition to giving sensible priors for the likelihood optimization, Eq.~\eref{eq:logL_parameter_estimation}, altitude atmospheric characterization would provide correct forms for the water vapor column density function, Eq.~\eref{eq:chis}, the temperature, Eq.~\eref{eq:temp_adiab}, and the wind profile.

\subsection{Upper limit on the polarization fraction of atmosphere emission}
\label{ssec:atm_polarization}
\begin{figure}[htb!]
	\centering
		\includegraphics[width=\columnwidth]{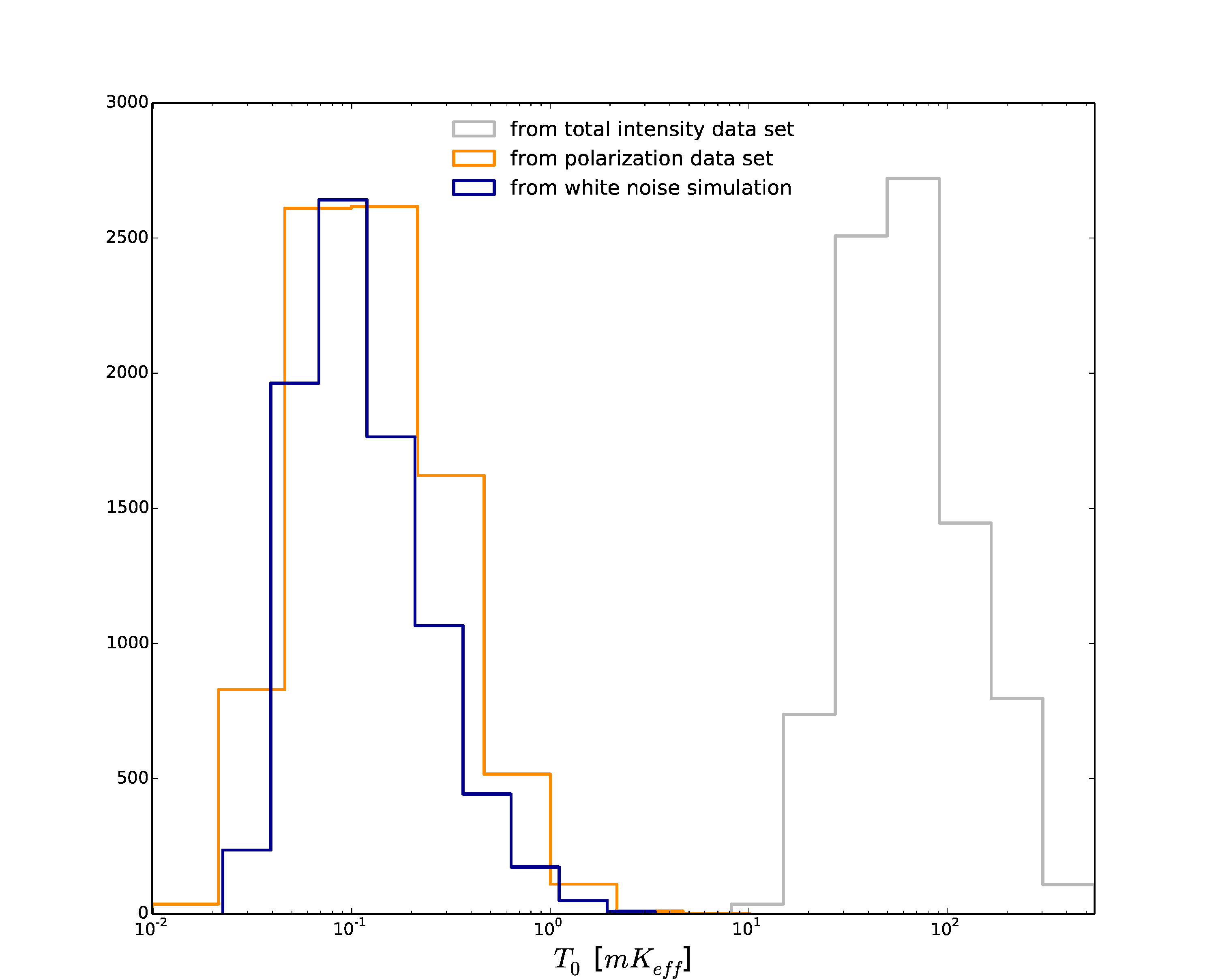}
		\caption{Distribution of the fitted amplitude of the correlations, $T_0$, obtained from the analysis of the total intensity (gray histogram, similar to the right panel of Fig.~\ref{fig:atmosphere_detection}), polarized data set (orange histogram) and from white noise simulated TODs (blue histogram).}
	\label{fig:T0_distribution_polarization}
\end{figure}

We perform the same analysis described in section~\ref{ssec:dijt_from_real_data}, but with data correlation matrices $\mathbf{D}_{ij}^{\,tt'}$ estimated from pixel difference time streams instead of pixel sum. In such case, the nearly scan-synchronous features described earlier are highly suppressed and the global amplitude of the correlation is significantly lower. The absence of features in $\mathbf{D}_{ij}^{\;tt'}$ makes the optimization of the likelihood more challenging, and usually results with a large wind speed and a low $T_0$. Fig.~\ref{fig:T0_distribution_polarization} shows the distribution of the amplitude $T_0$ from the analysis of pixel sum time streams (i.e., total intensity data, similar to the right panel of Fig.~\ref{fig:atmosphere_detection}) and from the pixel difference time streams (i.e., polarized data). We compare these results with the estimation of $T_0$ from simulated white noise TOD, taken as random Gaussian realizations assuming a distribution width equals to the standard deviation of the real time streams.\\
This simple approach shows that the amplitude of the polarized atmospheric signal is at least two orders of magnitude below the total intensity signal, from $60\pm30\,mK_{\rm eff}$ down to $\leq 300 \,\mu K_{\rm eff}$. Moreover, we verify that this latter value is compatible with the analysis of simulated white noise time streams. This indicates that constraints on $T_0$ from polarized data set mainly come from the uncorrelated detector noise. 
Hence, from the analysis of the \textsc{polarbear-i} first season of observation, the linear polarization fraction of the atmosphere emission $p$ satisfies $ p\,\leq\,1.0\%$.

\section{Conclusions}
\label{sec:conclusion}

We present a \com{new} modeling for the atmospheric emission in the millimeter and sub-millimeter wavelengths, and fit it to the \textsc{Polarbear-i} data sets.  \com{This work introduces the equations to evaluate the six-dimensional spatial correlation induced by turbulent water vapor. This correlation is a key object to realistically simulate a three-dimensional atmospheric medium, and also simulate the TOD of a detector associated to a given scanning strategy. By comparing the model predictions with \textsc{Polarbear-i} data sets, we derived distributions of the main atmospheric parameters such as the typical size of the turbulence, wind speed, temperature, etc. The semi-analytical equations, along with the typical distributions of physical parameters, are the ingredients to perform realistic simulations of atmosphere. This ability might be crucial to evaluate and optimize the performances of future ground-based CMB experiments.} \\

We first review in section~\ref{sec:photon-noise} the atmospheric transmission and background emission in the range of frequencies involved in CMB observations. We describe the thermal loading induced by atmosphere as well as the frequency and PWV dependence of the average contamination. 

In addition to the loading due to the atmosphere, the fluctuations present in this turbulent medium induce a correlated contamination for the detectors time streams of a given focal plane. On the grounds of the modeling originally introduced by~\citet{1995MNRAS.272..551C}, we derive in section~\ref{sec:atmosphere_physical_model} a \com{new expression} for both the auto- and cross-correlation induced by atmospheric fluctuations between two detectors of a given focal plane geometry. 

We present in section~\ref{sec:numerical_computation} several results from \com{an original} numerical computation of the model, performed using Quasi Monte Carlo algorithms. The computation of the modeling involves a six-dimensional integral for a given pair of detectors and a given time sample: numerical approximations of the integrals are required, and turn out to give sensible results. 
The main prediction of the modeling is the presence of nearly scan-synchronous features in the correlation function, modulated by the scanning strategy, the wind speed and direction, the atmospheric fluctuations typical size and the atmosphere temperature. These features are however only significant for a limited set of atmosphere parameters, and may not be relevant for the estimation of astrophysical or cosmological quantities.

We compare, in section~\ref{sec:atmosphere_estimation}, the prediction of the modeling with real CMB data sets from the \textsc{polarbear-i} first season \com{of observation}, using a maximum parametric likelihood approach. We derive estimates of four atmosphere parameters, $\{$$L_o$, $T_0$, $W$, $\phi_W$$\}$, and show that their distribution over the year were peaked around $\{$$300\pm 100$ m, $60\pm30\,mK_{\rm eff}$, $25\pm10$ m s$^{-1}$, $0\,\deg$$\}$ respectively.
The weakest parameter estimation concerns the wind direction $\phi_W$, which mainly impacts the relative position in time of the features present in the cross-correlations. The most likely value, $\phi_W = 0\,\deg$, is interpreted as an artifact of the analysis: it corresponds to a bound imposed on the parameter during the likelihood optimization. We do not see a significant improvement by cutting out cases with high reduced $\chi^2$.\\
We compare our estimate of $T_0$ with the brightness of atmospheric fluctuations measured by \cite{2000ApJ...543..787L} above the Atacama desert. To perform this comparison, we use a semi-analytical 2d layer approximation of the angular correlation estimated from the 3d modeling of the turbulences. Measurements by \cite{2000ApJ...543..787L} and from this work are $1\sigma$ compatible. 
\com{Finally, we give an updated observational upper limit on the polarization fraction of the atmospheric emission at $150$ GHz:} from the analysis of the \textsc{polarbear-i} polarized data sets, we find that the \com{linear} polarization fraction is $\,\leq\,1.0\%$.\\

External information about wind, temperature and water vapor profiles as measured by e.g., a flying weather probe above the considered observatory would certainly help for the estimation of the atmospheric conditions from the CMB data sets. Measurements of turbulence characteristics performed by an independent instrument such as MASS/DIMM would certainly be useful too (\cite{2007MNRAS.382.1268K}). Adding priors to the likelihood equation, Eq.~\eref{eq:logL_parameter_estimation}, can potentially speed up its maximization and more robustly constrain the different modeling parameters. In addition, note that we treat here each \textsc{polarbear-i} CES independently for the parameters estimation, but one \com{could imagine grouping} several CES together (for which atmosphere properties would be assumed \com{to be stable}), possibly helping in estimating or constraining parameters.\\

Parametrization and estimation of $\mathbf{C}_{ij}^{\; tt'}$ can be of great potential for the analysis of ground-based experiment data sets. In particular, the proposed modeling gives a basis for the development of simulation tools. We propose to study in a following work the effect of systematic effects on the performance of CMB polarization observations, in the presence of realistic atmospheric contamination, detector noise and polarized sky signal.
Coming CMB experiments, such as Stage-IV instruments, will use hundreds of thousands detectors: understanding, characterizing and potentially treating correlated noise will be critical for a recovery of their full sensitivity and an optimized exploitation of their data sets. 
In particular, we note that having multichromatic detectors, and potentially atmosphere monitoring detectors outside the atmospheric windows (although still in the optically thin limit), can help data analysts to better estimate the atmospheric signal, using for example frameworks similar to the astrophysical foregrounds separation techniques, e.g., \cite{2008A&A...491..597L}. We also leave the implementation and testing of this possibility to a future work.

\acknowledgments{ Calculations were performed on the National Energy Research Scientific Computing (NERSC), supported by the Department of Energy under Contract No. DE-AC02-05CH11231. The \textsc{Polarbear} project is funded by the National Science Foundation under grants AST-0618398 and AST-1212230.
The James Ax Observatory operates in the Parque Astronomico Atacama in Northern Chile under the auspices of the Comision Nacional de Investigacion Cientifica y Tecnologica de Chile (CONICYT). Finally, we would like to acknowledge the tremendous contributions by Huan Tran to the \textsc{Polarbear} instrument and who has been at the initiation of this particular project on atmosphere characterization.}

\bibliographystyle{apj}
\bibliography{references}

\begin{thebibliography}{}
\expandafter\ifx\csname natexlab\endcsname\relax\def\natexlab#1{#1}\fi

\bibitem[{{Abazajian} {et~al.}(2013){Abazajian}, {Arnold}, {Austermann},
  {Benson}, {Bischoff}, {Bock}, {Bond}, {Borrill}, {Calabrese}, {Carlstrom},
  {Carvalho}, {Chang}, {Chiang}, {Church}, {Cooray}, {Crawford}, {Dawson},
  {Das}, {Devlin}, {Dobbs}, {Dodelson}, {Dore}, {Dunkley}, {Errard}, {Fraisse},
  {Gallicchio}, {Halverson}, {Hanany}, {Hildebrandt}, {Hincks}, {Hlozek},
  {Holder}, {Holzapfel}, {Honscheid}, {Hu}, {Hubmayr}, {Irwin}, {Jones},
  {Kamionkowski}, {Keating}, {Keisler}, {Knox}, {Komatsu}, {Kovac}, {Kuo},
  {Lawrence}, {Lee}, {Leitch}, {Linder}, {Lubin}, {McMahon}, {Miller},
  {Newburgh}, {Niemack}, {Nguyen}, {Nguyen}, {Page}, {Pryke}, {Reichardt},
  {Ruhl}, {Sehgal}, {Seljak}, {Sievers}, {Silverstein}, {Slosar}, {Smith},
  {Spergel}, {Staggs}, {Stark}, {Stompor}, {Vieregg}, {Wang}, {Watson},
  {Wollack}, {Wu}, {Yoon}, \& {Zahn}}]{2013arXiv1309.5383A}
{Abazajian}, K.~N., {Arnold}, K., {Austermann}, J., {et~al.} 2013, ArXiv
  e-prints, arXiv:1309.5383

\bibitem[{{Ade} {et~al.}(2014){Ade}, {Akiba}, {Anthony}, {Arnold}, {Atlas},
  {Barron}, {Boettger}, {Borrill}, {Borys}, {Chapman}, {Chinone}, {Dobbs},
  {Elleflot}, {Errard}, {Fabbian}, {Feng}, {Flanigan}, {Gilbert}, {Grainger},
  {Halverson}, {Hasegawa}, {Hattori}, {Hazumi}, {Holzapfel}, {Hori}, {Howard},
  {Hyland}, {Inoue}, {Jaehnig}, {Jaffe}, {Keating}, {Kermish}, {Keskitalo},
  {Kisner}, {Le Jeune}, {Lee}, {Leitch}, {Linder}, {Lungu}, {Matsuda},
  {Matsumura}, {Meng}, {Miller}, {Morii}, {Moyerman}, {Myers}, {Navaroli},
  {Nishino}, {Paar}, {Peloton}, {Poletti}, {Quealy}, {Rebeiz}, {Reichardt},
  {Richards}, {Ross}, {Rotermund}, {Schanning}, {Schenck}, {Sherwin},
  {Shimizu}, {Shimmin}, {Shimon}, {Siritanasak}, {Smecher}, {Spieler},
  {Stebor}, {Steinbach}, {Stompor}, {Suzuki}, {Takakura}, {Tikhomirov},
  {Tomaru}, {Wilson}, {Yadav}, {Zahn}, \& {Polarbear
  Collaboration}}]{2014PhRvL.112m1302A}
{Ade}, P.~A.~R., {Akiba}, Y., {Anthony}, A.~E., {et~al.} 2014, Physical Review
  Letters, 112, 131302

\bibitem[{{Arnold}(2010)}]{2010PhDT.......176A}
{Arnold}, K.~S. 2010, PhD thesis, University of California, Berkeley

\bibitem[{{BICEP2 Collaboration} {et~al.}(2014){BICEP2 Collaboration}, {Ade},
  {Aikin}, {Barkats}, {Benton}, {Bischoff}, {Bock}, {Brevik}, {Buder},
  {Bullock}, {Dowell}, {Duband}, {Filippini}, {Fliescher}, {Golwala},
  {Halpern}, {Hasselfield}, {Hildebrandt}, {Hilton}, {Hristov}, {Irwin},
  {Karkare}, {Kaufman}, {Keating}, {Kernasovskiy}, {Kovac}, {Kuo}, {Leitch},
  {Lueker}, {Mason}, {Netterfield}, {Nguyen}, {O'Brient}, {Ogburn}, {Orlando},
  {Pryke}, {Reintsema}, {Richter}, {Schwarz}, {Sheehy}, {Staniszewski},
  {Sudiwala}, {Teply}, {Tolan}, {Turner}, {Vieregg}, {Wong}, \&
  {Yoon}}]{2014arXiv1403.3985B}
{BICEP2 Collaboration}, {Ade}, P.~A.~R., {Aikin}, R.~W., {et~al.} 2014, ArXiv
  e-prints, arXiv:1403.3985

\bibitem[{{Bussmann} {et~al.}(2005){Bussmann}, {Holzapfel}, \&
  {Kuo}}]{2005ApJ...622.1343B}
{Bussmann}, R.~S., {Holzapfel}, W.~L., \& {Kuo}, C.~L. 2005, \apj, 622, 1343

\bibitem[{{Chiang} {et~al.}(2010){Chiang}, {Ade}, {Barkats}, {Battle},
  {Bierman}, {Bock}, {Dowell}, {Duband}, {Hivon}, {Holzapfel}, {Hristov},
  {Jones}, {Keating}, {Kovac}, {Kuo}, {Lange}, {Leitch}, {Mason}, {Matsumura},
  {Nguyen}, {Ponthieu}, {Pryke}, {Richter}, {Rocha}, {Sheehy}, {Takahashi},
  {Tolan}, \& {Yoon}}]{2010ApJ...711.1123C}
{Chiang}, H.~C., {Ade}, P.~A.~R., {Barkats}, D., {et~al.} 2010, \apj, 711, 1123

\bibitem[{{Church}(1995)}]{1995MNRAS.272..551C}
{Church}, S.~E. 1995, \mnras, 272, 551

\bibitem[{{Crites} {et~al.}(2014){Crites}, {Henning}, {Ade}, {Aird},
  {Austermann}, {Beall}, {Bender}, {Benson}, {Bleem}, {Carstrom}, {Chang},
  {Chiang}, {Cho}, {Citron}, {Crawford}, {De Haan}, {Dobbs}, {Everett},
  {Gallicchio}, {Gao}, {George}, {Gilbert}, {Halverson}, {Hanson},
  {Harrington}, {Hilton}, {Holder}, {Holzapfel}, {Hoover}, {Hou}, {Hrubes},
  {Huang}, {Hubmayr}, {Irwin}, {Keisler}, {Knox}, {Lee}, {Leitch}, {Li},
  {Liang}, {Luong-Van}, {McMahon}, {Mehl}, {Meyer}, {Mocanu}, {Montroy},
  {Natoli}, {Nibarger}, {Novosad}, {Padin}, {Pryke}, {Reichardt}, {Ruhl},
  {Saliwanchik}, {Sayre}, {Schaffer}, {Smecher}, {Stark}, {Story}, {Tucker},
  {Vanderlinde}, {Vieira}, {Wang}, {Whitehorn}, {Yefremenko}, \&
  {Zahn}}]{2014arXiv1411.1042C}
{Crites}, A.~T., {Henning}, J.~W., {Ade}, P.~A.~R., {et~al.} 2014, ArXiv
  e-prints, arXiv:1411.1042

\bibitem[{{Das} {et~al.}(2013){Das}, {Louis}, {Nolta}, {Addison},
  {Battistelli}, {Bond}, {Calabrese}, {Devlin}, {Dicker}, {Dunkley},
  {D{\"u}nner}, {Fowler}, {Gralla}, {Hajian}, {Halpern}, {Hasselfield},
  {Hilton}, {Hincks}, {Hlozek}, {Huffenberger}, {Hughes}, {Irwin}, {Kosowsky},
  {Lupton}, {Marriage}, {Marsden}, {Menanteau}, {Moodley}, {Niemack}, {Page},
  {Partridge}, {Reese}, {Schmitt}, {Sehgal}, {Sherwin}, {Sievers}, {Spergel},
  {Staggs}, {Swetz}, {Switzer}, {Thornton}, {Trac}, \&
  {Wollack}}]{2013arXiv1301.1037D}
{Das}, S., {Louis}, T., {Nolta}, M.~R., {et~al.} 2013, ArXiv e-prints,
  arXiv:1301.1037

\bibitem[{{Delabrouille} {et~al.}(2003){Delabrouille}, {Cardoso}, \&
  {Patanchon}}]{2003MNRAS.346.1089D}
{Delabrouille}, J., {Cardoso}, J.-F., \& {Patanchon}, G. 2003, \mnras, 346,
  1089

\bibitem[{{Errard}(2012)}]{thesis_jojo}
{Errard}, J. 2012, PhD thesis, Universit\'e Paris VII Denis Diderot,
  http://tel.archives-ouvertes.fr/tel-00761117

\bibitem[{{Giovanelli} {et~al.}(2001){Giovanelli}, {Darling}, {Henderson},
  {Hoffman}, {Barry}, {Cordes}, {Eikenberry}, {Gull}, {Keller}, {Smith}, \&
  {Stacey}}]{2001PASP..113..803G}
{Giovanelli}, R., {Darling}, J., {Henderson}, C., {et~al.} 2001, \pasp, 113,
  803

\bibitem[{{Hanany} \& {Rosenkranz}(2003)}]{hanany_rosenkraz}
{Hanany}, S., \& {Rosenkranz}, P. 2003, \nar, 47, 1159

\bibitem[{{Hobgood}(1993)}]{1993IJCli..13..461H}
{Hobgood}, J.~S. 1993, International Journal of Climatology, 13, 461

\bibitem[{{Kasten} \& {Young}(1989)}]{1989ApOpt..28.4735K}
{Kasten}, F., \& {Young}, A.~T. 1989, \ao, 28, 4735

\bibitem[{{Kermish} {et~al.}(2012){Kermish}, {Ade}, {Anthony}, {Arnold},
  {Aubin}, {Boettger}, {Borrill}, {Cantalupo}, {Dobbs}, {Errard}, {Flanigan},
  {Ghribi}, {Halverson}, {Hazumi}, {Holzapfel}, {Howard}, {Hyland}, {Jaffe},
  {Keating}, {Kisner}, {Lee}, {Linder}, {Lungu}, {Matsumura}, {Miller}, {Meng},
  {Myers}, {Nishino}, {O'Brient}, {O'Dea}, {Reichardt}, {Schanning}, {Shimizu},
  {Shimmin}, {Shimon}, {Spieler}, {Steinbach}, {Stompor}, {Suzuki}, {Tomaru},
  {Tran}, {Tucker}, {Quealy}, {Richards}, \& {Zahn}}]{ziggy_SPIE}
{Kermish}, Z., {Ade}, P.~A.~R., {Anthony}, A.~E., {et~al.} 2012, in SPIE
  proceedings

\bibitem[{{Kogut} {et~al.}(2011){Kogut}, {Fixsen}, {Chuss}, {Dotson}, {Dwek},
  {Halpern}, {Hinshaw}, {Meyer}, {Moseley}, {Seiffert}, {Spergel}, \&
  {Wollack}}]{2011arXiv1105.2044K}
{Kogut}, A., {Fixsen}, D.~J., {Chuss}, D.~T., {et~al.} 2011, ArXiv e-prints,
  arXiv:1105.2044

\bibitem[{{Kornilov} {et~al.}(2007){Kornilov}, {Tokovinin}, {Shatsky},
  {Voziakova}, {Potanin}, \& {Safonov}}]{2007MNRAS.382.1268K}
{Kornilov}, V., {Tokovinin}, A., {Shatsky}, N., {et~al.} 2007, \mnras, 382,
  1268

\bibitem[{{Lay}(1997)}]{1997A&AS..122..535L}
{Lay}, O.~P. 1997, \aaps, 122, 535

\bibitem[{{Lay} \& {Halverson}(2000)}]{2000ApJ...543..787L}
{Lay}, O.~P., \& {Halverson}, N.~W. 2000, \apj, 543, 787

\bibitem[{{Leach} {et~al.}(2008){Leach}, {Cardoso}, {Baccigalupi}, \& {et
  al}}]{2008A&A...491..597L}
{Leach}, S.~M., {Cardoso}, J., {Baccigalupi}, C., \& {et al}. 2008, \aap, 491,
  597

\bibitem[{{Marriage}(2006)}]{thesis_marriage}
{Marriage}, T. 2006, PhD thesis, Princeton University

\bibitem[{{Masciadri} {et~al.}(2013){Masciadri}, {Lascaux}, \&
  {Fini}}]{2013MNRAS.436.1968M}
{Masciadri}, E., {Lascaux}, F., \& {Fini}, L. 2013, \mnras, 436, 1968

\bibitem[{{Naess} {et~al.}(2014){Naess}, {Hasselfield}, {McMahon}, {Niemack},
  {Addison}, {Ade}, {Allison}, {Amiri}, {Battaglia}, {Beall}, {de Bernardis},
  {Bond}, {Britton}, {Calabrese}, {Cho}, {Coughlin}, {Crichton}, {Das},
  {Datta}, {Devlin}, {Dicker}, {Dunkley}, {D{\"u}nner}, {Fowler}, {Fox},
  {Gallardo}, {Grace}, {Gralla}, {Hajian}, {Halpern}, {Henderson}, {Hill},
  {Hilton}, {Hilton}, {Hincks}, {Hlozek}, {Ho}, {Hubmayr}, {Huffenberger},
  {Hughes}, {Infante}, {Irwin}, {Jackson}, {Muya Kasanda}, {Klein}, {Koopman},
  {Kosowsky}, {Li}, {Louis}, {Lungu}, {Madhavacheril}, {Marriage}, {Maurin},
  {Menanteau}, {Moodley}, {Munson}, {Newburgh}, {Nibarger}, {Nolta}, {Page},
  {Pappas}, {Partridge}, {Rojas}, {Schmitt}, {Sehgal}, {Sherwin}, {Sievers},
  {Simon}, {Spergel}, {Staggs}, {Switzer}, {Thornton}, {Trac}, {Tucker},
  {Uehara}, {Van Engelen}, {Ward}, \& {Wollack}}]{2014JCAP...10..007N}
{Naess}, S., {Hasselfield}, M., {McMahon}, J., {et~al.} 2014, \jcap, 10, 7

\bibitem[{{Nash}(1984)}]{1984SJNA...21..770N}
{Nash}, S.~G. 1984, SIAM Journal on Numerical Analysis, 21, 770

\bibitem[{{Pardo} {et~al.}(2001){Pardo}, {Cernicharo}, \&
  {Serabyn}}]{2001ITAP...49.1683P}
{Pardo}, J.~R., {Cernicharo}, J., \& {Serabyn}, E. 2001, IEEE Transactions on
  Antennas and Propagation, 49, 1683

\bibitem[{{Pham} \& {Cardoso}(2001)}]{2001ITSP...49.1837P}
{Pham}, D.-T., \& {Cardoso}, J.-F. 2001, IEEE Transactions on Signal
  Processing, 49, 1837

\bibitem[{{Planck Collaboration} {et~al.}(2013){Planck Collaboration}, {Ade},
  {Aghanim}, {Alves}, {Armitage-Caplan}, {Arnaud}, {Ashdown},
  {Atrio-Barandela}, {Aumont}, {Aussel}, \& et~al.}]{2013arXiv1303.5062P}
{Planck Collaboration}, {Ade}, P.~A.~R., {Aghanim}, N., {et~al.} 2013, ArXiv
  e-prints, arXiv:1303.5062

\bibitem[{{POLARBEAR Collaboration} {et~al.}(2013){POLARBEAR Collaboration},
  {Ade}, {Akiba}, {Anthony}, {Arnold}, {Atlas}, {Barron}, {Boettger},
  {Borrill}, {Chapman}, {Chinone}, {Dobbs}, {Elleflot}, {Errard}, {Fabbian},
  {Feng}, {Flanigan}, {Gilbert}, {Grainger}, {Halverson}, {Hasegawa},
  {Hattori}, {Hazumi}, {Holzapfel}, {Hori}, {Howard}, {Hyland}, {Inoue},
  {Jaehnig}, {Jaffe}, {Keating}, {Kermish}, {Keskitalo}, {Kisner}, {Le Jeune},
  {Lee}, {Linder}, {Leitch}, {Lungu}, {Matsuda}, {Matsumura}, {Meng}, {Miller},
  {Morii}, {Moyerman}, {Myers}, {Navaroli}, {Nishino}, {Paar}, {Peloton},
  {Quealy}, {Rebeiz}, {Reichardt}, {Richards}, {Ross}, {Schanning}, {Schenck},
  {Sherwin}, {Shimizu}, {Shimmin}, {Shimon}, {Siritanasak}, {Smecher},
  {Spieler}, {Stebor}, {Steinbach}, {Stompor}, {Suzuki}, {Takakura}, {Tomaru},
  {Wilson}, {Yadav}, \& {Zahn}}]{2013arXiv1312.6646P}
{POLARBEAR Collaboration}, {Ade}, P.~A.~R., {Akiba}, Y., {et~al.} 2013, ArXiv
  e-prints, arXiv:1312.6646

\bibitem[{{Sayers} {et~al.}(2010){Sayers}, {Golwala}, {Ade}, {Aguirre}, {Bock},
  {Edgington}, {Glenn}, {Goldin}, {Haig}, {Lange}, {Laurent}, {Mauskopf},
  {Nguyen}, {Rossinot}, \& {Schlaerth}}]{2010ApJ...708.1674S}
{Sayers}, J., {Golwala}, S.~R., {Ade}, P.~A.~R., {et~al.} 2010, \apj, 708, 1674

\bibitem[{{Shimon} {et~al.}(2008){Shimon}, {Keating}, {Ponthieu}, \&
  {Hivon}}]{2008PhRvD..77h3003S}
{Shimon}, M., {Keating}, B., {Ponthieu}, N., \& {Hivon}, E. 2008, \prd, 77,
  083003

\bibitem[{{Spinelli} {et~al.}(2011){Spinelli}, {Fabbian}, {Tartari}, {Zannoni},
  \& {Gervasi}}]{2011MNRAS.414.3272S}
{Spinelli}, S., {Fabbian}, G., {Tartari}, A., {Zannoni}, M., \& {Gervasi}, M.
  2011, \mnras, 414, 3272

\bibitem[{{Stompor} {et~al.}(2009){Stompor}, {Leach}, {Stivoli}, \&
  {Baccigalupi}}]{2009MNRAS.392..216S}
{Stompor}, R., {Leach}, S., {Stivoli}, F., \& {Baccigalupi}, C. 2009, \mnras,
  392, 216

\bibitem[{{Tatarskii}(1961)}]{1961wptm.book.....T}
{Tatarskii}, V.~I. 1961, {Wave Propagation in Turbulent Medium} (McGraw-Hill)

\bibitem[{{Tauber} {et~al.}(2010){Tauber}, {Norgaard-Nielsen}, {Ade}, {Amiri
  Parian}, {Banos}, {Bersanelli}, {Burigana}, {Chamballu}, {de Chambure},
  {Christensen}, {Corre}, {Cozzani}, {Crill}, {Crone}, {D'Arcangelo},
  {Daddato}, {Doyle}, {Dubruel}, {Forma}, {Hills}, {Huffenberger}, {Jaffe},
  {Jessen}, {Kletzkine}, {Lamarre}, {Leahy}, {Longval}, {de Maagt}, {Maffei},
  {Mandolesi}, {Mart{\'{\i}}-Canales}, {Mart{\'{\i}}n-Polegre}, {Martin},
  {Mendes}, {Murphy}, {Nielsen}, {Noviello}, {Paquay}, {Peacocke}, {Ponthieu},
  {Pontoppidan}, {Ristorcelli}, {Riti}, {Rolo}, {Rosset}, {Sandri}, {Savini},
  {Sudiwala}, {Tristram}, {Valenziano}, {van der Vorst}, {van't Klooster},
  {Villa}, \& {Yurchenko}}]{2010A&A...520A...2T}
{Tauber}, J.~A., {Norgaard-Nielsen}, H.~U., {Ade}, P.~A.~R., {et~al.} 2010,
  \aap, 520, A2

\bibitem[{{Taylor}(1938)}]{1938RSPSA.164..476T}
{Taylor}, G.~I. 1938, Royal Society of London Proceedings Series A, 164, 476

\bibitem[{{The POLARBEAR Collaboration} {et~al.}(2014){The POLARBEAR
  Collaboration}, {Ade}, {Akiba}, {Anthony}, {Arnold}, {Atlas}, {Barron},
  {Boettger}, {Borrill}, {Chapman}, {Chinone}, {Dobbs}, {Elleflot}, {Errard},
  {Fabbian}, {Feng}, {Flanigan}, {Gilbert}, {Grainger}, {Halverson},
  {Hasegawa}, {Hattori}, {Hazumi}, {Holzapfel}, {Hori}, {Howard}, {Hyland},
  {Inoue}, {Jaehnig}, {Jaffe}, {Keating}, {Kermish}, {Keskitalo}, {Kisner}, {Le
  Jeune}, {Lee}, {Leitch}, {Linder}, {Lungu}, {Matsuda}, {Matsumura}, {Meng},
  {Miller}, {Morii}, {Moyerman}, {Myers}, {Navaroli}, {Nishino}, {Paar},
  {Peloton}, {Poletti}, {Quealy}, {Rebeiz}, {Reichardt}, {Richards}, {Ross},
  {Schanning}, {Schenck}, {Sherwin}, {Shimizu}, {Shimmin}, {Shimon},
  {Siritanasak}, {Smecher}, {Spieler}, {Stebor}, {Steinbach}, {Stompor},
  {Suzuki}, {Takakura}, {Tomaru}, {Wilson}, {Yadav}, \&
  {Zahn}}]{2014arXiv1403.2369T}
{The POLARBEAR Collaboration}, {Ade}, P.~A.~R., {Akiba}, Y., {et~al.} 2014,
  ArXiv e-prints, arXiv:1403.2369

\bibitem[{{Wright} \& {Nocedal}(2006)}]{optimization_book}
{Wright}, S., \& {Nocedal}, J. 2006, {Numerical Optimization} (Springer)

\bibitem[{{Wu} {et~al.}(2014){Wu}, {Errard}, {Dvorkin}, {Kuo}, {Lee},
  {McDonald}, {Slosar}, \& {Zahn}}]{2014ApJ...788..138W}
{Wu}, W.~L.~K., {Errard}, J., {Dvorkin}, C., {et~al.} 2014, \apj, 788, 138

\end{thebibliography}

\begin{appendix}
\vspace{1cm}
\section{A: Semi-analytical comparison with the frozen 2d screen approximation}
\label{sec:3d_and_2d_discussion}

The approximation of a 2d frozen screen assumes an angular correlation given by (\cite{2000ApJ...543..787L})
\begin{eqnarray}
	\centering
		C_{\rm 2d\ screen}(\gamma) = 2\pi\int_0^{\infty}{d\psi\, \psi\, P(\psi)\, J_0 (2\pi\psi \gamma})
	\label{eq:C_theta_bussmann}
\end{eqnarray}
where $\psi$ is the angular wave number, $P(\psi)$ is the angular power spectrum and $J_0$ is the $0th$-order Bessel function. 
It is assumed that the observed field $\gamma$ is small and that the turbulent atmospheric layer is high i.e.,
\begin{eqnarray}
	\centering
		\Delta h \gg \frac{h}{2\sin({\theta})}\, \gamma.
	\label{eq:Dh_h_theta}
\end{eqnarray}
This layer is located at an altitude $h$ with a thickness $\Delta h$, and the observation is made at an elevation $\theta$. In the case of the Kolmogorov turbulence power law, Eq.~\eref{eq:Pk_kolm},  \cite{2005ApJ...622.1343B} gives an expression for $P(\psi)$ of the form
\begin{eqnarray}
	\centering
		P(\psi) = B_\nu^2\sin({\theta})^{-8/3}\,\psi^{-11/3},
	\label{eq:C_theta_bussmann_2} 
\end{eqnarray}
where the small-angle approximation ($\gamma \ll 1$) is employed.

We show in the following derivation that the 3d modeling is general enough to recover the angular correlation of the 2d screen approach.
To compare Eq.~\eref{eq:C_theta_bussmann} with the correlation $\mathbf{C}_{ij}^{\;tt'}$ introduced in Eq.~\eref{eq:auto2}, we can collapse the 3d modeling into a thick 2d layer. We assume for simplification that the lines-of-sight for detectors $i$ and $j$ are pointing at the zenith, i.e., $\mathbf{r^i_s},\,\mathbf{r^j_s}\, \propto\, \mathbf{\hat{z}}$. Similarly to \cite{1995MNRAS.272..551C}, Eq.~\eref{eq:auto2} can be expressed as
\begin{eqnarray}
	\centering
		C^{h\pm\frac{\Delta h}{2}}_{\rm 3d\ model}(\gamma) \simeq \sqrt{2\pi}L_o \int{ dZ\, \chi_2(Z)\, \left(T_{\rm phys}(Z)\right)^2 } \times\left( 1 + \frac{w^2(Z,\, \gamma)}{2L_o^2} \right)^{-1}
	\label{eq:auto2_approx}
\end{eqnarray} 
where $Z\equiv (z + z')/2$, and we assumed that $w^2(z) \simeq w^2(z')$ which is only valid if $|z-z'| \ll Z$. We also use the notation $\chi_2(Z) = \chi_2(z,z')$. Eq.~\eref{eq:auto2_approx} is derived by expressing Eq.~\eref{eq:auto2} in cartesian coordinates, and by changing the integration variables from $\{x,y,z,x',y',z'\}$ to $\{(x+x')/2,(y+y')/2,(z+z')/2,x-x',y-y',z-z'\}$. 
\begin{figure}
	\centering
		\includegraphics[width=10cm]{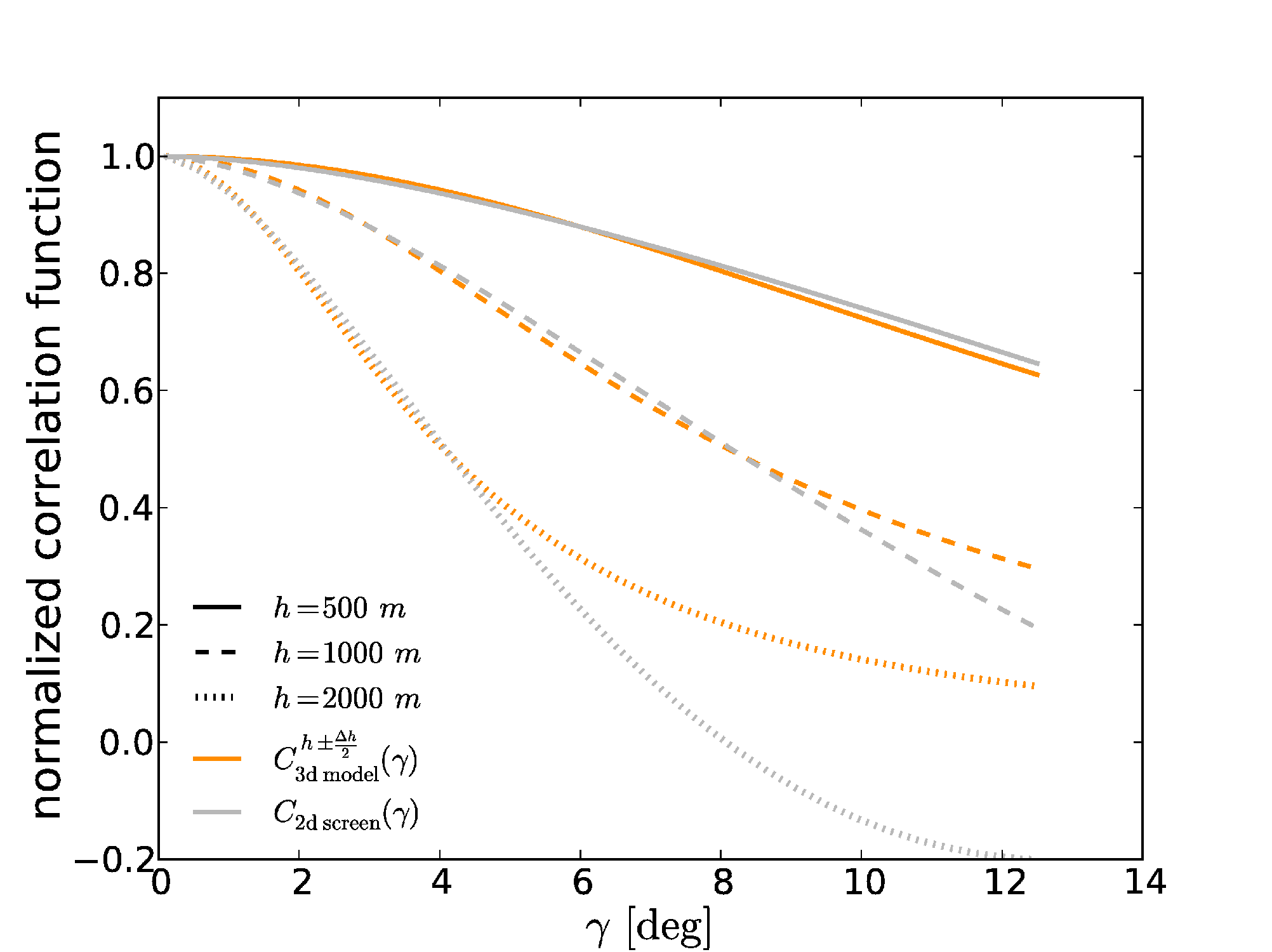}
	\caption{Comparison of the normalized angular correlations: $C_{\rm 2d\ screen}(\gamma)$ from Eq.~\eref{eq:C_theta_bussmann} and $C^{h\pm\frac{\Delta h}{2}}_{\rm 3d\ model}(\gamma)$ from Eq.~\eref{eq:comparison_2d_3d} as a function of the angular scale $\gamma$. We assume $\Delta h = 500$ m, ${\rm el} = 90\,\deg$ and $L_o=100$ m. The lower bound for the integral in Eq.~\eref{eq:C_theta_bussmann} is taken as $h/(2\Delta h)$.}
	\label{fig:C_vs_theta_comparison}
\end{figure}
The integral over $Z$ can be limited to a turbulent 2d layer by taking a step function for $\chi_2$ such as
\begin{eqnarray}
	\centering
		\chi_2 (z,z') &=& 1.0\, {\rm m}^{-2}\ {\rm for\ \{z,z'\}\, \in\, \left[ h\pm\frac{\Delta h}{2} \right]}.
	\label{eq:chi2_approx}
\end{eqnarray}
We can hence derive an approximation of the collapsed 2d correlation $C^{h\pm\frac{\Delta h}{2}}_{\rm 3d\ model}(\gamma)$ from Eq.~\eref{eq:auto2_approx},
\begin{eqnarray}
		C^{h\pm\frac{\Delta h}{2}}_{\rm 3d\ model}(\gamma) \simeq\sqrt{2\pi}L_o\, T_0^2 \left( 1 + \frac{(w(h,\, \gamma))^2}{2L_o^2} \right)^{-1} \times\,[1.0\,m^{-2}]\times\, \Delta h \,\left( 1 - \frac{h}{z_{\rm atm}} \right)^2,
	\label{eq:comparison_2d_3d}
\end{eqnarray}
where we took a constant Gaussian beam width $w(Z)\simeq w(h)\simeq h\, \gamma$, and assumed that $\Delta h \ll z_{\rm atm}$.\\

Eq.~\eref{eq:comparison_2d_3d} shows that larger $L_o$ leads to larger correlation for a fixed angular scale $\gamma$. A thicker ($\Delta h\nearrow$) or a warmer ($T_0\nearrow$, $z_{\rm atm}\nearrow$) atmospheric layer would increase the correlation between measurements separated by an angle $\gamma$. Moreover, larger angular separation $\gamma$ would lead to smaller correlation for fixed $L_o$, $T_0$, $h$ and $\Delta h$. Fig.~\ref{fig:C_vs_theta_comparison} shows the dependence of the correlation functions $C_{\rm 2d\ screen}(\gamma)$ from Eq.~\eref{eq:C_theta_bussmann} and $C^{h\pm\frac{\Delta h}{2}}_{\rm 3d\ model}(\gamma)$ from Eq.~\eref{eq:comparison_2d_3d}, as a function of the angular scale $\gamma $. We look at various $h \,\in\,\{500\,{\rm m},\,1000\,{\rm m},\,2000\,{\rm m}\}$, take $L_o=100$ m and $\Delta h =500$ m.  We consider a lower bound $h/(2 \Delta h)$ for the integral in Eq.~\eref{eq:C_theta_bussmann}, since this sets the range of $\psi$ for which Kolmogorov spectrum is valid (\cite{2000ApJ...543..787L}).
In all the considered cases, the fractional difference between $C^{h\pm\frac{\Delta h}{2}}_{\rm 3d\ model}(\gamma)$ and  $C_{\rm 2d\ screen}(\gamma)$ is $\leq\,10\%$ for $\gamma\leq\,5\,\deg$.\\

This shows that the 3d modeling is general enough to reproduce the 2d screen correlation function.  In fact, some parameters such as $\{h, \Delta h, L_o, z_{\rm atm}\}$ can be tuned so that $C_{\rm 2d\ screen} \simeq C^{h\pm\frac{\Delta h}{2}}_{\rm 3d\ model}$ on small angular scales. 

\vspace{1cm}
\section{B: Approximation of the full noise covariance matrix}
\label{sec:approx_cov}

We propose an algorithm to build an approximation of $\mathbf{D_{ij}^{\; tt'}}$, less expensive computationally, lighter on disk, easier to handle. In the approximation that  $\mathbf{D_{ij}^{\; tt'}}$ is stationary, its Fourier transform is expected to be diagonal. We therefore build the full binned covariance matrix, named  $\mathbf{D_{ij}^{\; b}}$ in the following: $i$ and $j$ are individual detectors (or equivalently sum and differences of perpendicular detectors within a focal plane pixel) and $b$ is a given frequency bin. 
However, note that the stationarity assumption is broken in reality and the formalism below could not be used to recover e.g., the wind properties from real data sets.\\
We first introduce the algebra and its implementation, and second propose a framework to generate TOD from $\mathbf{D}_{ij}^{\; b}$. 

\subsection{Binned noise covariance estimation from TOD}
\label{ssec:Cijb_introduction}

Similarly to what was proposed in~\cite{2010ApJ...711.1123C}, each considered time stream is apodized and Fourier transformed:
\begin{eqnarray}
	\centering
		\forall\ t\,\in\, S,\  d_i^{\ (f,\ S)} &\equiv& {\rm FFT} \left( \bar{W}_{t}\times d_i^{\ t} \right)
\end{eqnarray}
where $S$ denotes a group of consecutive subscans and $\bar{W_t}$ is the normalization of a hanning window, $W_t$, such that
 \begin{equation}
 \centering
 	\bar{W_t} \equiv \frac{W_t}{\left\langle W_t^2 \right\rangle_t}.
\end{equation}
Windowing prevents aliasing between frequency modes, which would bias the evaluation of the noise, especially at low frequencies. 
To describe the structure of the correlations between detectors as a function of frequency, we choose a set of $n_{\rm bins}$ frequency bins $b$, which should be optimized to encapsulate noise properties (scan synchronous contaminations, low frequency power, high frequency effects, etc.). We describe below a suitable choice for this set, in the case of \textsc{polarbear}.
For a given CES, the full binned covariance matrix is defined as
\begin{eqnarray}
	\centering
		\forall\ {i,j,b},\hspace{0.2cm} \mathbf{D}_{ij}^{\; b} \equiv \left\langle \left\langle  d_{i}^{\ (f,\ S)} d_{j}^{\;\star\ (f,\ S)}  \right\rangle_{f\,\in\,b} \right\rangle_{S}
	\label{eq:Cijb_def}
\end{eqnarray}
$\mathbf{D}_{ij}^{\; b}$ is a complex object of size $n_{\rm det} \times n_{\rm det} \times n_{\rm bins}$, which satisfies the symmetry relation
\begin{eqnarray}
	\centering
		\forall\ {i,j,b},\hspace{0.2cm} \mathbf{D}_{ij}^{\; b} = \left( \mathbf{D}_{ji}^{\; b}\right)^\star
	\label{eq:Cijb_symmetry}
\end{eqnarray}

\begin{figure*}[htb!]
	\centering
	\includegraphics[width=16cm]{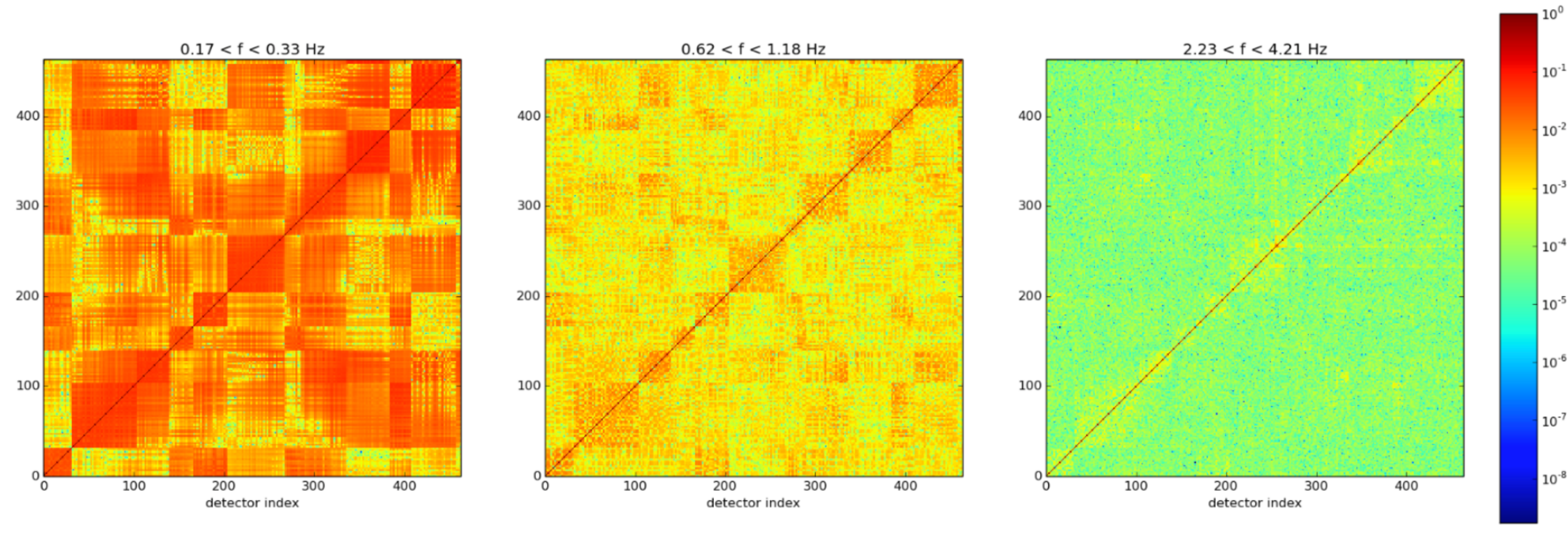}
	\caption{\footnotesize Three normalized $\mathbf{D}_{ij}^{\; b}$ shown as 2d objects, for different frequency bins $b$. The $x$ and $y$ axis correspond to individual detector indices: in this case, $i$ and $j$ correspond to individual bolometers. The applied normalization is the following: $\forall\, i,j,b,\ \mathbf{D}_{ij}^{\; b} \rightarrow\,\mathbf{D}_{ij}^{\; b}/\sqrt{\mathbf{D}_{ii}^{\; b}\times \mathbf{D}_{jj}^{\; b}}$.  The depicted quantity has therefore no units and the matrices diagonals are equal to $1$. One can notice that the correlations between different detectors (off diagonal elements) decrease at higher frequencies. In the case of \textsc{polarbear-i}, polarization noise turned out to be fairly white in the chosen science bands, for $500 < \ell < 2000$. The squared patterns noticeable in the left panel correspond to common structures of the focal plane: detectors are physically close and share some parts of the readout, leading to expected correlation.}
	\label{fig:Cijb_illustration}
\end{figure*}

In Fig.~\ref{fig:Cijb_illustration} are depicted several typical $\mathbf{D}_{ij}^{\; b}$, shown here as 2-dimensional objects (in the $\{$det $i$, det $j \}$ space), for three different bins $b$. In this example, $i$ and $j$ indices correspond to single detectors, i.e. \textsc{polarbear-i} bolometers. 
Because of the stationarity assumption, $\mathbf{D}_{ij}^{\; b}$ misses some information about atmospheric contamination. However, one can observe the detector-detector correlations, which can be quite significant at low frequencies, $f \leq 1$ Hz.\\

To avoid mode coupling and to better estimate the low frequency part of the noise, we consider groups of 25 consecutive subscans for the analysis. Note that $S$ can be composed of an odd number of subscans since the phase information (left- and right-going scans are not equivalent because of the wind) is anyway destroyed during the construction of $\mathbf{D}_{ij}^{\; b}$, based on the stationarity assumption.\\
In addition, similarly to the computation of $\mathbf{D}_{ij}^{\; t t'}$ in Eq.~\eref{eq:Dijtau_def}, accelerating or decelerating motions of the telescope (turnarounds) are also included, as they did not show to have significantly different noise properties. We use the symmetry relation, Eq.~\eref{eq:Cijb_symmetry}, to speed up the computation time.

Besides, bins are chosen to be logarithmically spaced in the  $\left[ 0.01,\,15 \right]$ Hz range. An interesting possibility is to use an adaptive binning, which would make sure that any important feature in the noise power spectra is optimally encapsulated by the covariance matrix.

\subsection{TOD simulation given $\mathbf{D}_{ij}^{\; b}$}
\label{ssec:simulation_from_cijb}

In this paragraph we propose a procedure to simulate realistic TODs using the informations encapsulated in the $\mathbf{D}_{ij}^{\; b}$ quantity. 
We follow the usual method to simulate noise-only TOD from the full binned noise covariance matrix, $\mathbf{D}_{ij}^{\; b}$.
First, we define
\begin{eqnarray}
	\centering
		\forall\ b,\ \mathbf{L}_{ij}^{\; b} \equiv \sqrt{ \mathbf{D}_{ij}^{\; b} }
	\label{eq:Lijb_def}
\end{eqnarray}
where $\sqrt{\cdot}$ is any operation such that, for any $\mathbf{A}$ which has a square root, 
\begin{eqnarray}
	\centering
		\sqrt{ \mathbf{A}}\,\sqrt{ \mathbf{A}}^{\ \dagger} = \mathbf{A}
\end{eqnarray}
For each frequency $f$, we generate a $n_{\rm det}$ random vector $\xi_f\ \in \ \mathbb{C}^{n_{\rm det}}$.
We can write 
\begin{eqnarray}
	\centering
		\xi_f \equiv \xi_f^{\; r} + i\,\xi_f^{\; i}
\end{eqnarray}
where $\xi_f^{\; r}$ and $\xi_f^{\; i} \in \mathbb{R}$ are independent random variables.
The simulated noise time stream, $n_{i}^{\; f}$, expressed in Fourier domain, will be given by
\begin{eqnarray}
	\centering
		 \forall\ f\,\in\,\left\{ f_{\rm min}, f_{\rm max} \right\} ,\ \ n_{i}^{\; f} = \sum_{j} \mathbf{L}_{ij}^{\; b}\, \xi_{j}^{\; f},
	\label{eq:noise_generation}
\end{eqnarray}
in which equation one should remember that $b = b(f)$. In addition, $f_{\rm min} = 1/T$, $T$ being the length of the simulated TOD, and $f_{\rm max}$ is the maximum frequency achievable.
In our implementation we choose $T=$ (length of a CES), corresponding to $f_{\rm min} \leq {\rm min }(b)$. Therefore, we decide to apply a same power $\mathbf{L}_{ij}^{\; b=0}$ to the lowest frequency modes of the TODs.
Of course, one would like have a first bin $b$ as low as possible in frequency. However, one has to take into account the average over $S$ used in Eq.~\eref{eq:Cijb_def}: taking $S \sim 25$ consecutive subscans, leading to $\sim 5-10$ groups of $S$ in the case of \textsc{polarbear-i} scanning strategy, turned out to give enough informations about the lowest frequency modes of the TODs.

It is possible to interpolate $\mathbf{L}_{ij}^{\; b}$ over the frequency bins, leading to a $\mathbf{L}_{ij}^{\; f}$ matrix. Finally, the time domain simulated noise, for a detector $i$, is therefore given by
\begin{eqnarray}
	\centering
		n_{i}^{\ t} = {\rm FFT}^{-1}\left(n_{i}^{\; f}\right).
	\label{eq:nit_fft_inv_nif}
\end{eqnarray}
In practice, to implement Eq.~\eref{eq:Lijb_def}, we diagonalize $\mathbf{D}_{ij}^{\; b}$:
\begin{eqnarray}
	\centering
		\forall\ b,\ \mathbf{D}_{ij}^{\; b} = \sum_k e_k\; \mathbf{v_k}  \otimes \mathbf{v_k}^\dagger,
\end{eqnarray}
where $e_k$ and $\mathbf{v_k}$ are respectively the eigen values and eigen vectors.
We then simply write
\begin{eqnarray}
	\centering
		\forall\ b,\ \mathbf{L}_{ij}^{\; b}  = \sum_k \sqrt{e_k}\; \mathbf{v_k}  \otimes \mathbf{v_k}^\dagger.
\end{eqnarray}
Besides, our implementation uses a simple parallelization scheme, each process doing the TODs-simulation of a given CES, i.e., Eqs.~\eref{eq:noise_generation} and~\eref{eq:nit_fft_inv_nif}. All the $\mathbf{D}_{ij}^{\; b}$ and their square root $\mathbf{L}_{ij}^{\; b}$ (Eq.~\eref{eq:Lijb_def}), are precomputed and saved on disk for the entire season.

\subsection{Discussion}

For a given frequency bin $b$, the matrix $\mathbf{L}_{ij}^{\; b}$ has a rank $R = n_{f\in b} \times n_{S}$, where $n_{f\in b}$ is the number of frequencies included in the bin $b$ and $n_{S}$ is the number of groups of subscans. It turns out that, for a given $b$, $\mathbf{L}_{ij}^{\; b}$ is not full rank i.e., $R < n_{\rm det}^{\; 2}$. Consequently, there are $n_{\rm det}^{\; 2} - R$ singular eigen values, making $\mathbf{D}_{ij}^{\; b}$ semi-positive definite for a given $b$---and preventing for example its Cholesky decomposition. We checked that this singularity did not have any impact on the simulations.

\end{appendix}

\vspace{0.5cm}
\noindent\rule{6cm}{0.4pt}
\vspace{0.5cm}

\footnotesize{
\altaffilmark{1}{AstroParticule et Cosmologie, Univ Paris Diderot, CNRS/IN2P3, CEA/Irfu, Obs de Paris, Sorbonne Paris Cit\'e, France}

\altaffilmark{2}{Center for Astrophysics and Space Astronomy, University of Colorado, Boulder, CO 80309, USA}

\altaffilmark{3}{Computational Cosmology Center, Lawrence Berkeley National Laboratory, Berkeley, CA 94720, USA}

\altaffilmark{4}{Department of Astronomy and Astrophysics, University of Chicago, Chicago, IL 60637, USA}

\altaffilmark{5}{Department of Astronomy, Pontifica Universidad Catolica de Chile}

\altaffilmark{6}{Department of Astrophysical and Planetary Sciences, University of Colorado, Boulder, CO 80309, USA}

\altaffilmark{7}{Department of Electrical and Computer Engineering, University of California, San Diego, CA 92093, USA}

\altaffilmark{8}{Department of Physics and Astronomy, University of Pennsylvania, Philadelphia, PA 19104-6396, USA}

\altaffilmark{9}{Department of Physics and Atmospheric Science, Dalhousie University, Halifax, NS, B3H 4R2, Canada}

\altaffilmark{10}{Department of Physics, Columbia University, New York, NY 10027, USA}

\altaffilmark{11}{Department of Physics, Imperial College London, London SW7 2AZ, United Kingdom}

\altaffilmark{12}{Department of Physics, Princeton University, Princeton, NJ 08544, USA}

\altaffilmark{13}{Department of Physics, University of California, Berkeley, CA 94720, USA}

\altaffilmark{14}{Department of Physics, University of California, San Diego, CA 92093-0424, USA}

\altaffilmark{15}{Department of Physics, University of Colorado, Boulder, CO 80309, USA}

\altaffilmark{16}{High Energy Accelerator Research Organization (KEK), Tsukuba, Ibaraki 305-0801, Japan}

\altaffilmark{17}{International School for Advanced Studies (SISSA), Trieste 34014, Italy}

\altaffilmark{18}{Kavli Institute for Cosmological Physics, University of Chicago, Chicago, IL 60637, USA}

\altaffilmark{19}{Kavli Institute for the Physics and Mathematics of the Universe (WPI), Todai Institutes for Advanced Study, The University of Tokyo, Kashiwa, Chiba 277-8583, Japan}

\altaffilmark{20}{Miller Institute for Basic Research in Science, University of California, Berkeley, CA 94720, USA}

\altaffilmark{21}{Observational Cosmology Laboratory, Code 665, NASA Goddard Space Flight Center, Greenbelt, MD 20771, USA}

\altaffilmark{22}{Osaka University, Toyonaka, Osaka 560-0043, Japan}

\altaffilmark{23}{Physics Department, Austin College, Sherman, TX 75090, USA}

\altaffilmark{24}{Physics Department, McGill University, Montreal, QC H3A 0G4, Canada}

\altaffilmark{25}{Physics Division, Lawrence Berkeley National Laboratory, Berkeley, CA 94720, USA}

\altaffilmark{26}{Rutherford Appleton Laboratory, STFC, Harwell, Oxford OX11 0QX, United Kingdom}

\altaffilmark{27}{School of Physics and Astronomy, Cardiff University, Cardiff CF10 3XQ, United Kingdom}

\altaffilmark{28}{School of Physics and Astronomy, Tel Aviv University, Tel Aviv 69978, Israel}

\altaffilmark{29}{School of Physics, University of Melbourne, Parkville, VIC 3010, Australia}

\altaffilmark{30}{Space Sciences Laboratory, University of California, Berkeley, CA 94720, USA}

\altaffilmark{31}{The Graduate University for Advanced Studies, Hayama, Miura District, Kanagawa 240-0115, Japan}

\altaffilmark{32}{Department of Physics and Astronomy, University of California, Irvine, USA}

\altaffilmark{33}{JAXA, Japan}

}

\end{document}